# Excitation functions of proton-induced reactions on natural Nd in the 10-30 MeV energy range, and production of radionuclides relevant for double-β decay

O. Lebeda[1,*], V. Lozza[2], P. Schrock[2†], J. Štursa[1] and K. Zuber[2]

[1]*Nuclear Physics Institute of the Academy of Sciences of the Czech Republic, Husinec-Řež, Czech Republic*
[2]*Technische Universität Dresden, Institut für Kern-und Teilchenphysik, Dresden, Germany*

A preferred candidate for neutrinoless double-β decay, $^{150}$Nd, is present in natural neodymium at an abundance level of 5.64%. However, neodymium could be activated by cosmic rays during the period it spends on the Earth's surface. Its activation by protons is therefore of interest when it comes to estimating the possible disturbance effects and increased background during neutrinoless double-β decay experiments like Sudbury Neutrino Observatory plus liquid scintillator (SNO+). In most cases, we lack experimental data on the proton-induced reactions on neodymium. Therefore, a measurement of cross sections has been performed for the formation of $^{141}$Pm, $^{143}$Pm, $^{144}$Pm, $^{146}$Pm, $^{148}$Pm, $^{148}$Pm$^{m}$, $^{149}$Pm, $^{150}$Pm, $^{140}$Nd, $^{141}$Nd, $^{147}$Nd, $^{149}$Nd, $^{138}$Pr$^{m}$, $^{139}$Pr, $^{142}$Pr, and $^{139}$Ce by 10–30 MeV protons. Oxidation-protected metal foil targets of natural isotopic abundance were irradiated by the usual stacked-foil technique on the external proton beam of the isochronous cyclotron U-120M at the Nuclear Physics Institute at Řež near Prague. Special attention was paid to the excitation functions of long-lived radionuclides. The measured data were compared with TENDL-2010 library data (TALYS code).

## I. INTRODUCTION

The experimental observation of double-β decay (i.e., a change in nuclear charge by two units while leaving the atomic mass constant) plays a key role in solving the problem concerning the unknown absolute neutrino mass value and the properties of neutrinos under CP-conjugation. Double β-decay (ββ-decay) is a second-order weak decay, observable for those even-even nuclei where β decay is energetically forbidden, or at least strongly suppressed. In addition to the standard-model process resulting in the emission of two electrons and two antineutrinos, there is the neutrinoless double-β-decay mode (ββ0ν) which violates the total lepton number by two units. This decay requires that neutrinos have a nonvanishing rest mass to match helicities and that they are Majorana particles. This would be physics beyond the standard model and other lepton-number-violating physics could contribute as well [1]. The decay rate is defined, following the Fermi golden rule, as:

$$\left(T_{1/2}^{0\nu}\right)^{-1} = G^{0\nu}(Q_{\beta\beta}, Z)\left|M^{0\nu}\right|^2 \left\langle m_\nu \right\rangle^2,$$

where G is the phase space integral, M is the nuclear matrix element and $\left\langle m_\nu \right\rangle^2$ the so-called effective Majorana neutrino mass. From all potential 35 double-β emitters, a very suitable candidate is $^{150}$Nd. It has the second-highest Q value of 3371.38 ± 0.20 keV [2] and also a reasonable nuclear matrix elements; for a recent compilation see [3]. Currently, three experiments are planning to study the neutrinoless double-β decay of $^{150}$Nd; namely, the large mass Nd-loaded scintillator experiment called Sudbury Neutrino Observatory plus liquid scintillator (SNO+) [4], Drift Chamber Beta-ray Analyzer DCBA [5], and SuperNEMO [6], the last two using Nd in the form of foils within Time Projection Chambers (TPCs).

SNO+ is the follow up of the SNO experiment that will search – among other physical goals – for the neutrinoless double-β decay. In a first double-β phase, the liquid scintillator will be loaded with 0.1% of $^{nat}$Nd that contains $^{150}$Nd as double-β emitter. Since the expected signal is very small, much attention must be paid to the purity of the scintillator material in order to decrease the background. One of the Nd-related background components is due to the decay of radionuclides produced in $^{nat}$Nd by cosmic rays. Neodymium could get activated on the Earth's surface while being transported to the underground lab. Among the most important reactions are those due to neutrons and protons which can produce long-lived radionuclides with *Q* values up to 3.5 MeV; high enough to fall directly or pile up in the energy range of interest for the neutrinoless double-β-decay study. The hadron component of the cosmic flux at the Earth's surface is composed mainly of neutrons (95%), protons (3%), and π mesons (2%) [7,8]. Even if the proton flux is usually less than 3% of the total hadron flux [9], it may result in comparable or even more relevant activation of $^{nat}$Nd with respect to the SNO+ than that caused by the neutron flux. This is due to the fact that potential proton activation products are significantly longer lived than potential neutron activation products. While (*p*,*xn*) reactions result often in radionuclides with half-lives of the order of a year (see below), the longest-lived product of the neutron radiative capture is $^{147}$Nd (10.98 d), the longest-lived product of the (*n*,*p*) reactions is $^{143}$Pr (13.57 d) and only the less-probable (*n*,*α*) and (*n*,*αxn*) reactions can produce long-lived $^{139}$Ce (137.6 d), $^{141}$Ce (32.5 d), and $^{144}$Ce (284.9 d). Therefore, this article is devoted solely to the measurement of the elemental production cross sections of protons on natural neodymium.

* lebeda@ujf.cas.cz
† Present address: Institut für Kernphysik, Technical University Darmstadt, Darmstadt, Germany

Proton-induced reactions on natural neodymium result in a large number of radionuclides, still at relatively low proton energy. Excitation functions for most of them have not been measured yet. Moreover, $^{nat}$Nd has 7 stable or long-lived naturally occurring isotopes – see Table I [10], and thus many of the radionuclides are produced by two or more simultaneous reactions on several of them. A review of the main radionuclides produced directly or indirectly by 10-30 MeV protons in natural neodymium together with the relevant reaction or decay channels is displayed in Table II. The measurement of the excitation functions provides necessary data for estimating the background in SNO+, the maximum allowed time on the surface for $^{nat}$Nd, and the cooling time before it could be added to the liquid scintillator. Moreover, the cross sections for $^{nat}$Nd(*p*,*x*) reactions are mostly missing in the current database. Reliable experimental data are necessary to check the forecast of the nuclear reaction model codes that might significantly differ from reality. The measured cross sections are naturally of interest for testing such model codes; in particular, in cases when they can be converted to a single reaction on a single isotope.

## II. EXPERIMENTAL SETUP

The excitation functions were measured by the usual activation technique on the external beam of the isochronous cyclotron U-120M in Řež. The cyclotron provides proton beams with variable energy up to 38 MeV. Seventeen natural neodymium targets of 10 × 10 mm$^2$ area were prepared from commercially available Nd foils (99.9% purity, AlfaAesar). The foil thickness was measured using low-energy gamma ray ($^{241}$Am source) absorption and was equal to 97.7±0.3 μm (i.e., 68.39 mg/cm$^2$). Since metallic Nd is one of the most reactive lanthanides, special attention was paid to its protection against oxidizing in air. We decided in favour of coating the metal foils by very thin polymers that provide good protection against air and water vapours, and at the same time allow for irradiating the foils with acceptably high beam currents. Two plastic materials fulfilled those requirements: parylene and polyethylene. The first can resist proton currents higher than 0.6 μA, while the second can withstand beam currents up to ca 0.3 μA (no visible damage to the coating). The parylene coating's better resistance and its lower thickness (only 2–3 μm) makes it more suitable for these irradiations. We thus used this coating in the experiment. The targets were fixed in a water-cooled holder designed for irradiating the foil stacks. Each stack consisted of one titanium and copper entrance monitor and the neodymium foils interleaved with copper degraders that served as beam monitors, too.

In order to avoid higher uncertainties in the energy in the foils, we irradiated stacks containing a maximum three Nd foils. The typical stack arrangement is given below:

(i) Titanium (thickness 12.11 μm) acting as a beam monitor,
(ii) Copper foil (thickness 10.6 μm) acting as a beam monitor,
(iii) Plastic-coated neodymium foil (thickness 97.7 μm),
(iv) Copper beam energy degrader (55.0 μm), inserted if necessary,
(v) Copper foil (Cu thickness 10.6 μm) as a beam monitor,
(vi) Plastic-coated neodymium foil (thickness 97.7 μm),
(vii) Copper-beam energy degrader (55.0 μm), inserted if necessary,
(viii) Copper foil (Cu thickness 10.6 μm) as a beam monitor,
(ix) Plastic-coated neodymium foil (thickness 97.7 μm),
(x) Copper foil (Cu thickness 10.6 μm) as a beam monitor,
(xi) Thick silver foil acting as a beam stop, directly cooled by water.

## III. RADIOACTIVITY MEASUREMENT

Radionuclides produced in the Nd targets and the titanium and copper monitors were identified and their activity was measured using a gamma spectrometer equipped with a high-purity Ge (HPGe) detector [Ortec GMX45Plus with 55% NaI(Tl) efficiency at 1332.5 keV]. The spectrometer was calibrated by a set of standards supplied by the Czech Institute of Metrology (combined standard uncertainties of their activities are given in brackets): $^{241}$Am (0.3%), $^{133}$Ba (0.6%), $^{60}$Co (1.0%) and $^{152}$Eu (0.5%). Detection efficiencies were calibrated at the distances 200, 400, 600, 1000 and 1600 mm to allow for the measurement of each sample at acceptable dead time as it decays. Logarithms of the measured efficiency data points were fitted against logarithms of energy both by a of 5th order polynomial and by a linear fit (for gamma-ray energies >240 keV). Both fits agreed very well with each other and their correlation coefficients were >0.999.

The foils were dismounted and the activity measurement was started ca25–35 min after the end of bombardment (EOB). All the spectra were collected by Maestro32 software and analysed independently by the groups from Technische Universität Dresden and the Nuclear Physics Institute at Řež [11,12].

In Table III, we summarize the half-lives and energies of the main gamma lines together with their intensities used for calculation of the activities of the Nd activation products and the products of the monitoring reactions. Data were taken from [13–15]. Each Nd foil was measured five times in order to determine activities of both the short- and long-lived radionuclides with a good precision. All the measurements were always performed at a distancesuch that the detector dead time was < 40 %, preferably < 20 % (however, the response of the gamma spectrometer was found to be linear up to 70 % of the dead time).

## IV. BEAM ENERGY AND BEAM CURRENT MEASUREMENTS

The beam energy was calculated from the precisely measured beam orbit position [16] and its decrease in the various materials was calculated using the code SRIM [17].

The energy value can be cross checked in the case of two simultaneously running monitoring reactions by a method described in [18]. The energies calculated from the precisely measured beam orbit position and the energies based on the ratio of activities of zinc radionuclides born in the copper monitors differed max. for ca. 1.5 MeV. However, the latter method is insensitive to the change of energy in some energy regions (e.g., 23–26 MeV), and the results can be thus burdened with relatively high uncertainty. Since the estimated uncertainty of the beam energy calculated from the precisely measured beam orbit position is only 0.2 MeV, we adopted that value as more reliable.

The beam current was measured using the activity of the monitoring reaction products and the well-known activation formula – cf. Equation (4). For that purpose, we used the beam current based on the $^{nat}Ti(p,x)^{48}V$ monitoring reaction. The activity of long-lived $^{48}V$ can be precisely determined including the fraction ejected by the beam to the following copper monitor that acted as a catcher foil, too (usually 1.5–2.0 % of the total $^{48}V$ activity produced was ejected).

## V. Data processing

To calculate the activity of a single radionuclide in a given foil at the EOB, we used only well-resolved peaks of the highest intensities available (see Table III). Since the Nd foils were relatively thick, the net peak area was corrected for the mean attenuation of a given gamma line in the foil. Attenuation coefficients were obtained from the X-COM [19]. After correcting the net peak area for the decay during the real time of the measurement, the activity at the EOB was calculated applying standard corrections for the gamma line's detection efficiency and intensity, for the time of measurement and the decay between the start of the measurement and the EOB.

In some cases, the measured activity had two components: a certain fraction of the radionuclide's nuclei was produced directly in the nuclear reaction(s), and the rest of nuclei was born indirectly as a decay product of another radionuclide created in the target. This was the case of $^{141}Nd$, which is decay product of $^{141}Pm$, and the case of $^{149}Pm$ which is decay product of $^{149}Nd$. If the content of both fractions is significant, one can correct the net peak area for contribution of the indirectly born nuclei, as described elsewhere [20]. Briefly, the activity of the daughter radionuclide at the EOB is given by the following formula:

$$A_2^{EOB} = \frac{f\left(1 - \frac{\lambda_2}{\lambda_2 - \lambda_1} e^{-\lambda_1 t_b} + \frac{\lambda_1}{\lambda_2 - \lambda_1} e^{-\lambda_2 t_b}\right)}{\left(1 - e^{-\lambda_1 t_b}\right)} A_1^{EOB} \qquad (1),$$

where $A_2^{EOB}$ is the activity of a daughter radionuclide at the EOB [Bq], $A_1^{EOB}$ is the activity of a parent radionuclide at the EOB [Bq], $f$ is the probability of transition of parent to daughter radionuclide, $\lambda_1$ is the decay constant of a parent radionuclide [$hr^{-1}$], $\lambda_2$ is the decay constant of a daughter radionuclide [$hr^{-1}$], and $t_b$ is time of irradiation (bombardment) [hr].

After the EOB, the activity of a daughter radionuclide can be described easily as follows:

$$A_2 = \frac{\lambda_2}{\lambda_2 - \lambda_1} f A_1^{EOB}\left(e^{-\lambda_1 t} - e^{-\lambda_2 t}\right) + A_2^{EOB} e^{-\lambda_2 t} \qquad (2),$$

where $t$ is the time passed between the EOB and the start of the activity measurement.

Correction of the net peak area for the contribution of indirectly born radionuclide is then given as number of counts corresponding to the mean activity $\bar{A}_2$ of a daughter radionuclide during measurement time $t_m$ derived from Eq. (2):

$$\overline{A}_2 = \frac{\lambda_2 f A_1^0}{(\lambda_2 - \lambda_1)t_m}\left(\frac{1-e^{-\lambda_1 t_m}}{\lambda_1} - \frac{1-e^{-\lambda_2 t_m}}{\lambda_2}\right) + \frac{A_2^0}{\lambda_2 t_m}\left(1-e^{-\lambda_2 t_m}\right), \qquad (3),$$

where $A_1^0$ and $A_2^0$ are the activities of the parent and daughter radionuclide at the start of the measurement. Activity $A_2^0$ is then calculated from Eq. (2).

In another cases, we could not directly measure the activity of the parent radionuclides contributing to the activity of their daughter products either due to the excessively short half-life or due to the missing or low-intensity gamma lines. This concerns the following decay chains:

$$^{140}\mathrm{Pm} \xrightarrow{9.2\ \mathrm{s,\ EC\ 100\%}} {}^{140}\mathrm{Nd} \xrightarrow{3.37\ \mathrm{d,\ EC\ 100\%}} {}^{140}\mathrm{Pr} \xrightarrow{3.39\ \mathrm{min,\ EC\ 100\%}} {}^{140}\mathrm{Ce}$$

$$^{140m}\mathrm{Pm} \xrightarrow{5.95\ \mathrm{min,\ EC\ 100\%}} {}^{140}\mathrm{Nd} \xrightarrow{3.37\ \mathrm{d,\ EC\,100\%}} {}^{140}\mathrm{Pr} \xrightarrow{3.39\ \mathrm{min,\ EC\ 100\%}} {}^{140}\mathrm{Ce}$$

$$^{139}\mathrm{Pr} \xrightarrow{4.41\ \mathrm{hr,\ EC\ 100\%}} {}^{139}\mathrm{Ce} \xrightarrow{137.641\,\mathrm{d,\ EC\ 100\%}} {}^{139}\mathrm{La}$$

$$^{147}\mathrm{Pr} \xrightarrow{13.4\ \mathrm{min},\ \beta^-\ 100\%} {}^{147}\mathrm{Nd} \xrightarrow{10.98\,\mathrm{d},\ \beta^-\ 100\%} {}^{147}\mathrm{Pm} \xrightarrow{2.6234\,\mathrm{a},\ \beta^-\ 100\%} {}^{147}\mathrm{Sm}$$

$$^{149}\mathrm{Pr} \xrightarrow{2.26\ \mathrm{min},\ \beta^-\ 100\%} {}^{149}\mathrm{Nd} \xrightarrow{1.728\,\mathrm{hr},\ \beta^-\ 100\%} {}^{149}\mathrm{Pm} \xrightarrow{53.08\,\mathrm{hr},\ \beta^-\ 100\%} {}^{149}\mathrm{Sm}$$

Since we do not know the initial ratio of the ground and metastable state of $^{140}$Pm (too short-lived for the direct measurement), we decided to calculate cumulative cross sections for $^{140}$Nd. Although the latter has no gamma lines, it is possible to deduce its activity from the activity of its daughter $^{140}$Pr extrapolated to the EOB (in equilibrium, both activities are practically equal).

Similarly, we did not detect $^{139}$Pr due to its very weak gamma lines, but we could measure its daughter $^{139}$Ce very precisely. From its activity, we deduced the activity of its parent at the EOB.

The activity of $^{147}$Pr was also not detected. Cross sections of $^{147}$Nd are, therefore, cumulative, although the contribution from the decay of $^{147}$Pr seems to be relatively low. The same is true for $^{149}$Pr that has such a short half-life that its detection is hardly possible.

Finally, we did not detect some radionuclides that are certainly produced in the proton-induced reactions ($^{145}$Pm, $^{147}$Pm) both due to their long half-lives and due to missing or extremely low-intensity gamma lines.

Cross sections were then calculated using well-known activation formula:

$$\sigma = \frac{A^{EOB} Aze}{d\rho N_A I(1-e^{-\lambda t_b})}, \qquad (4),$$

where $\sigma$ is the production cross section for a given radionuclide in the foil centre [cm$^2$], $A^{EOB}$ is the activity of a given radionuclide produced in a foil at the EOB [Bq], $A$ is the atomic weight of the foil metal [g/mol], $z$ is the beam particle charge (for protons $z$ = 1), $e$ is the electron charge (1.602177 × $10^{-19}$ C), $d$ is the foil thickness [cm], $\rho$ is the density of the foil metal [g/cm$^3$], $N_A$ is Avogadro's number [6.022137 × $10^{23}$ mol$^{-1}$], $I$ is the beam current [A], λ is the decay constant of the produced radionuclide [hr$^{-1}$], and $t_b$ is the irradiation (bombardment) time [hr].

The overall cross-section uncertainty was calculated as a square root of the sum of squares of the following sources of uncertainty:

(i) beam current measurement 10%,
(ii) net peak area 0.4–28% (mostly <2.0%),
(iii) detection efficiency <3%,
(iv) gamma lines intensities <3%,
(v) overall 11–28.6% (mostly <12.0%).

## VI. Results and discussion

All the measured cross sections are displayed in Tables IV and V and selected are plotted in Figs 1–14 along with the calculated excitation functions based on TENDL-2010 database [21]. For the calculations, we took into account all the reaction channels contributing for formation of a given radionuclide indicated in the Table II.

### A. $^{141}$Pm

Activity of the short-lived $^{141}$Pm (20.9 min) was calculated from its 1223.26 keV gamma line that was well visible in the first spectrum of each irradiated foil except for the last three. The measured cross sections are displayed in Fig. 1. Calculated excitation functions taken from the TENDL database provides the cross-section values that are almost by a factor of 2 higher than the measured ones in the energy region of 20–30 MeV. However, the general shape of both the calculated and measured curves is in good agreement.

### B. $^{143}$Pm

Activity of the long-lived $^{143}$Pm (265 d) was calculated from its single 741.98 keV gamma line that is present on all the spectra measured closer to the detector. Activity could be thus cross-checked in several independent measurements – agreement was excellent (within statistical error of the net peak area). Results are displayed in Fig. 2 together with the excitation function obtained from the TENDL database. Obviously, shapes of both curves correspond very well to each other, although the calculated data are, in general, some 20% higher than the measured data.

### C. $^{144}$Pm

The long-lived $^{144}$Pm (363 d) was measured via its prominent 618.01 keV gamma line well visible in all the spectra measured closer to the detector. Agreement between several measurements was excellent. The results together with the TENDL data are presented in Fig. 3. Calculated and measured data correspond very well up to ca. 19 MeV. For proton energies >19 MeV, calculated data are for 10–15 % higher than the experimental points, but this difference is practically within their uncertainties.

### D. $^{146}$Pm

The radionuclide $^{146}$Pm (5.53 a) is the longest-lived activation product that we measured. The data are based on the net peak area of its 453.88 keV gamma line from the last measurement of each foil that provided the best counting statistics. Experimental data displayed in Fig. 4 show good agreement with the results in the TENDL database; only for proton energies >16 MeV the latter are higher than the former by 15–30%. The general trend of both curves is the same.

### E. $^{148}$Pm and $^{148}$Pm$^{m}$

Activities of $^{148}$Pm (5.37 d) and of its isomer $^{148}$Pm$^{m}$ (41.29 d) were calculated from the gamma lines that accompany only the decay of one of them; that is, we used the 1465.12 keV gamma line for the former and the 629.99 keV gamma line for the latter. Since the probability of isomeric transition of $^{148}$Pm$^{m}$ to its ground state is only 4.2%, it was not necessary to correct the activity of the ground state for the contribution of the isomeric state using the spectra taken shortly after the EOB. Experimental cross sections are displayed in Fig. 5 and Fig. 6. Although the shape of the excitation functions predicted by TENDL corresponds well to the measured data, the difference in absolute values is significant for both $^{148}$Pm and $^{148}$Pm$^{m}$. In the case of the ground state, TENDL gives up to double the cross sections for proton energies >16 MeV, whereas, in the case of the isomer, experimental data are roughly twice the value the code predicts (for proton energies >20 MeV) and the experimental maximum seems to be shifted by ca. 2 MeV compared to the TENDL database.

### F. $^{149}$Pm

Longer-lived $^{149}$Pm (53.08 hr) is one of the two radionuclides that are born by a single reaction on a single target nucleus in $^{nat}$Nd: $^{150}$Nd*(p,2n)*$^{149}$Pm. However, its activity – measured easily via its single 285.95 keV gamma line – is in fact the cumulative activity of both directly born $^{149}$Pm and the $^{149}$Pm resulting from the decay of $^{149}$Nd. Since we were able to measure the activity of $^{149}$Nd very precisely, we could easily correct the activity of $^{149}$Pm for that contribution, as described in Sec. IV Data Processing. Table IV contains both corrected and cumulative cross sections. Figure 7 displays only the corrected data that can be directly compared to the TENDL database. The shapes of measured and predicted excitation functions agree very well; the experimental data are only slightly higher for proton energies 16–24 MeV, and slightly lower for proton energies > 24 MeV than the prediction.

### G. $^{150}$Pm

Relatively short-lived $^{150}$Pm (2.68 hr) is the second radionuclide – next to $^{149}$Pm – that is born in a single reaction on a single target nucleus: $^{150}$Nd*(p,n)*$^{150}$Pm. Its activity was calculated from its prominent 333.97 keV gamma line. Measured cross sections significantly differ from the TENDL database, as is obvious from Fig. 8. The general trend is similar in both cases, but experimental data do not show any small "second maximum" as the database suggests, and the cross-section values are for proton energies <12 MeV up to double what TENDL predicts. On the other hand, for proton energies > 20 MeV, TENDL predicts cross sections higher by up to ca 30% compared by measured values.

### H. $^{140}$Nd

As thoroughly discussed in the Sec. IV, Data Processing, we decided to calculate at least the cumulative cross sections for $^{140}$Nd (3.37 d), since both of their parents, $^{140}$Pm and $^{140m}$Pm, are too short-lived to be directly measured. Moreover, we had to make use of its equilibrium with its short-lived daughter $^{140}$Pr to measure its activity, since decay of $^{140}$Nd itself is not accompanied by emission of any gamma line. For that purpose we employed the 1596.1 keV gamma line of $^{140}$Pr. A few cross sections measured at the highest energies are displayed in Table V.

### I. $^{141}$Nd

Radionuclide $^{141}$Nd (2.49 hr) is born both in nuclear reactions and as a decay product of $^{141}$Pm. Its activity, calculated via its prominent 1126.80 keV gamma line, must be, therefore, corrected for the contribution of $^{141}$Pm. Corrected values were used for calculation of the cross sections that are displayed on Fig. 9. Compared to the excitation function taken from the TENDL database, one finds a strong discrepancy between them: the TENDL prediction is up to 10 times smaller than the measured values, although all the contributing reaction channels indicated in the Table II were taken into account in the TENDL database.

### J. $^{147}$Nd

The longer-lived $^{147}$Nd (10.98 d) is born predominantly in the *(p,pn)* or *(p,d)* reactions on $^{148}$Nd, although some contribution to its activity from the decay of $^{147}$Pr (13.4 min) cannot be excluded. The latter can be born in the $^{150}$Nd*(p,α)* and $^{148}$Nd*(p,2p)* reactions; however, we have not detected $^{147}$Pr in any of the spectra, even those measured as the first after the EOB (they were, however, taken at larger sample-detector distances, so that low activities could be invisible, although present). We have, therefore, decided to consider the measured activity of $^{147}$Nd and, consequently, cross sections based on it to be cumulative. The activity of $^{147}$Nd was calculated via its 531.02 keV gamma line. The measured cross sections together with the results in the TENDL database are displayed on Fig. 10. As is obvious, for the energies <22 MeV, experimental data are significantly higher than the database prediction, whereas for the energies >26 MeV, they are significantly lower. The discrepancy for the lower energies might be partly due to ignoring the above mentioned contribution of the $^{147}$Pr, since in the TENDL prediction we could take into account only the $^{148}$Nd*(p,pn)* and $^{148}$Nd*(p,d)* reactions.

### K. $^{149}$Nd

Activity of the short-lived $^{149}$Nd (1.728 hr) was calculated via its 211.31 keV gamma line and the results correspond well with those obtained via its further gamma lines of high intensity. The radionuclide can be born directly in the $^{150}$Nd*(p,pn)* and $^{150}$Nd*(p,d)* reactions that were both taken into account in the TENDL prediction and also indirectly as a product of the decay of the short-lived $^{149}$Pr (2.26 min). The latter can be born only in the $^{150}$Nd*(p,2p)* reaction. The measured cross sections – that are, therefore, considered to be cumulative – and the results of the TENDL database are displayed on Fig. 11. The shape of the measured curve, absolute cross-section values and the differences between TENDL and experiment are very similar to those for $^{147}$Nd (see above).

### L. $^{138}$Pr$^{m}$

Relatively short-lived $^{138}$Pr$^{m}$ (2.12 hr) can originate in the reactions *(p,αxn)* on $^{142–144}$Nd. Its activity we measured via its 1037.80 keV gamma line and the radionuclide was detected only in spectra of the foils irradiated with proton energy > 1 MeV. The experimental cross sections and calculations taken from the TENDL are plotted in Fig. 12. In spite of the general agreement in the trend, experimental data are approximately 60% of the values predicted by the TENDL database.

### M. $^{139}$Pr

Radionuclide $^{139}$Pr (4.41 hr) can be born in the $^{142}$Nd*(p,α)* and *(p,αxn)* reactions on $^{143–146}$Nd. Since it has very weak gamma lines, its activity must be calculated from the activity of its daughter, longer-lived $^{139}$Ce (137.641 d) using its 165.86 keV gamma line. Figure 13 shows comparison between the experimental data and excitation function taken from the TENDL database. In contrast to TENDL, the measured cross sections only smoothly increase with the energy without any prominent maximum at ca 23.5 MeV as predicted by the database.

### N. $^{142}$Pr

Activity of $^{142}$Pr (19.12 hr) was detected only in the foils irradiated with proton energy >18 MeV and calculated via its 1575.6 keV gamma line. The radionuclide can be formed in the following reactions: $^{143}$Nd*(p,2p),* $^{144}$Nd*(p,$^3$He),* $^{145}$Nd*(p,α)* and $^{146}$Nd*(p,αn).* The measured data and excitation function taken from the TENDL database are given in Fig. 14. As obvious, the experimental data are in general significantly lower than the database predicts (by a factor of two and even more).

### O. $^{139}$Ce

This longer-lived radionuclide (137.641 d) is produced only indirectly as the decay product of $^{139}$Pr and its cumulative activity was measured via its prominent 165.86 keV gamma line. Therefore, the calculated cumulative cross sections are displayed only in Table V.

## VII. Discussion

The cross sections for formation of radionuclides by proton-induced reactions on natural neodymium were measured in the energy range 10–29 MeV. In most cases, the published cross sections are the first measured data, and they can serve as a test for the theoretical prediction by model codes. Comparison between the experimental data and the prediction of the TENDL database typically shows good agreement both in shapes and absolute values; in particular for the *(p,xn)* reactions. Sometimes the discrepancy appears in the absolute values; for example, in the case of $^{148}$Pm$^{m}$ and $^{148}$Pm, where the measured ratio of the ground and metastable states is ca. 2:1, while the code predicts <1, or in the case of $^{141}$Pm, $^{150}$Pm, $^{141}$Nd, $^{138}$Pr$^{m}$, $^{139}$Pr, and $^{142}$Pr. In the case of $^{139}$Pr and $^{142}$Pr, there is also significant difference in the shape of experimental and TENDL-based excitation functions. The isomeric cross-section ratio for the $^{148}$Nd*(p,n)* reaction was measured for highly enriched target material by Aumann and Gückel [22]. There is very good agreement between our data and theirs up to ca. 17 MeV; then they start to differ. However, for energies greater than 17 MeV, production of $^{148}$Pm and $^{148}$Pm$^{m}$ through the $^{150}$Nd*(p,3n)* channel becoming significant in comparison with production via the $^{148}$Nd*(p,n)* reaction that, in contrast, decreases with energy (cf. Fig. 15). Since in the present work an $^{nat}$Nd target was used, the contribution of both reaction channels has to be considered, and that explains the higher cross-sections ratio measured for $E_p$ > 17 MeV. In order to compare the data properly, the data of Aumann and Gückel were corrected for the different intensity of the 1465.12 keV gamma line of $^{148}$Pm (24.3%) used by these authors.

From the point of view of the SNO+ experiment, the most interesting radionuclides are the long-lived radionuclides with half-life >30 d ($^{143}$Pm, $^{144}$Pm, $^{146}$Pm, $^{148}$Pm$^{m}$, $^{139}$Ce). Physical thick target yields for these five radionuclides based on our experimental cross sections are displayed in Figs. 16 and 17. Since, except for $^{139}$Ce, we did not measure their excitation functions up to the threshold, we could calculate the yields for the leaving-proton energy equal to 10 MeV. The yields are, therefore, slightly underestimated, but they give a good idea about the production rates of these long-lived radionuclides relevant for SNO+. A precise calculation of the proton activation is not possible due to the unknown cosmic proton flux spectrum in this energy region. In Table VI we give a rough estimation of the expected activities of the long-lived radionuclides at sea level in the energy range from 10–30 MeV using the measured cross sections as well as the calculated cross sections. The proton density flux per energy unit is adopted from [8] and the amount of $^{nat}Nd$ taken into calculation is 1 ton. If highly enriched $^{150}Nd$ is considered, then the activities of $^{148}Pm$, $^{148}Pm^{m}$, and $^{147}Nd$ will increase for about 1/0.05638, since they are formed predominantly in the reactions with $^{150}Nd$.

## VIII. CONCLUSION

We have measured the production crosssections of different radionuclides in proton-induced reactions on natural neodymium in the 10–30 MeV energy range. The values for long-lived radionuclides differ by up to 100% with respect to the values calculated from the TENDL-2010 library data. We also calculated the corresponding production rates of long-lived radionuclides in the 10–30 MeV energy range in natural neodymium exposed to the cosmic proton flux. The values obtained give an idea about the content of radionuclidic impurities induced in natural neodymium by cosmic protons on the Earth's surface for estimated exposure time and cooling time underground prior to its use in the SNO+ experiment. The results confirmed that the predictive power of TALYS code is relatively good in the case of long-lived promethium radioisotopes in the energy range studied. This suggests that it can also be used in a wider energy range, at least for the preliminary estimation of activation by protons of higher incident energy.

## ACKNOWLEDGEMENTS

We thank Alex Fauler for the preparation of the parylene target coating, S. Belogurov for fruitful discussions, and U-120M cyclotron crew for irradiations. The work was supported by Deutsche Forschungsgemeinschaft (ZU 123/5-1) and by the Academy of Sciences of the Czech Republic under the NPI research plan AV0Z10480505.

TABLE I. Stable and semistable isotopes of neodymium and their abundances in natural element.

| Isotope | Abundance (%) | Half-life (a) | Decay mode |
|---|---|---|---|
| $^{142}$Nd | 27.152(40) | stable | – |
| $^{143}$Nd | 12.174(26) | stable | – |
| $^{144}$Nd | 23.798(19) | $2.29\times10^{15}$ | α |
| $^{145}$Nd | 8.293(12) | stable | – |
| $^{146}$Nd | 17.189(32) | stable | – |
| $^{148}$Nd | 5.756(21) | stable | – |
| $^{150}$Nd | 5.638(28) | $9.11\times10^{18}$ | $\beta^-\beta^-$ |

TABLE II. Activation products of the proton-induced reactions on natural neodymium together with contributing reaction channels and their Q values and with indirect decay paths. From all possible reaction channels on a given nucleus resulting in the same product, we give only that with maximum Q value (emission of the most complex particle). The Q values of the other channels can be then easily calculated by subtracting the respective binding energy from the maximum Q value (e.g., *d* = *np* + 2.225 MeV, *t* = *p2n* + 8.482 MeV, 3He = *2pn* + 7.718 MeV and *α* = *2p2n* + 28.296 MeV).

| Radionuclide | Reaction channels | Q value (MeV) |
|---|---|---|
| $^{140}$Pm+$^{140}$Pm$^{m}$ | $^{142}$Nd($p$,3$n$) | −24.674 |
| $^{141}$Pm | $^{142}$Nd($p$,2$n$) | −14.286 |
| | $^{143}$Nd($p$,3$n$) | −20.410 |
| $^{143}$Pm | $^{142}$Nd($p$,$\gamma$) | +4.300 |
| | $^{143}$Nd($p$,$n$) | −1.824 |
| | $^{144}$Nd($p$,2$n$) | −9.641 |
| | $^{145}$Nd($p$,3$n$) | −15.396 |
| | $^{146}$Nd($p$,4$n$) | −22.962 |
| $^{144}$Pm | $^{143}$Nd($p$,$\gamma$) | +4.703 |
| | $^{144}$Nd($p$,$n$) | −3.114 |
| | $^{145}$Nd($p$,2$n$) | −8.870 |
| | $^{146}$Nd($p$,3$n$) | −16.435 |
| $^{146}$Pm | $^{145}$Nd($p$,$\gamma$) | +5.312 |
| | $^{146}$Nd($p$,$n$) | −2.254 |
| | $^{148}$Nd($p$,3$n$) | −14.879 |
| $^{148}$Pm+$^{148}$Pm$^{m}$ | $^{148}$Nd($p$,$n$) | −1.324 |
| | $^{150}$Nd($p$,3$n$) | −13.744 |
| $^{149}$Pm | $^{148}$Nd($p$,$\gamma$) | +5.947 |
| | $^{150}$Nd($p$,2$n$) | −6.473 |
| | Decay of $^{149}$Nd | |
| $^{150}$Pm | $^{150}$Nd($p$,$n$) | −0.869 |
| $^{140}$Nd | Decay of $^{140}$Pm | |
| | $^{142}$Nd($p$,$t$) | −9.364 |
| | $^{143}$Nd($p$,$tn$) | −15.488 |
| | $^{144}$Nd($p$,$t$2$n$) | −23.305 |
| $^{141}$Nd+$^{141}$Nd$^{m}$ | $^{142}$Nd($p$,$d$) | −7.604 |
| | $^{143}$Nd($p$,$t$) | −7.470 |
| | $^{144}$Nd($p$,$tn$) | −15.287 |
| | $^{145}$Nd($p$,$t$2$n$) | −21.043 |
| | Decay of $^{141}$Pm | |
| $^{147}$Nd | $^{148}$Nd($p$,$d$) | −5.108 |
| | Decay of $^{147}$Pr | |
| $^{149}$Nd | $^{150}$Nd($p$,$d$) | −5.156 |
| | Decay of $^{149}$Pr | |
| $^{138}$Pr$^{m}$ | $^{142}$Nd(p,αn) | −6.031 |
| | $^{143}$Nd($p$,$\alpha$2$n$) | −12.155 |
| | $^{144}$Nd($p$,$\alpha$3$n$) | −19.972 |

TABLE II. (*Continued*.)

| Radionuclide | Reaction channels | Q value (MeV) |
|---|---|---|
| $^{139}$Pr | $^{142}$Nd($p$,$\alpha$) | +3.732 |
| | $^{143}$Nd($p$,$\alpha n$) | −2.391 |
| | $^{144}$Nd($p$,$\alpha 2n$) | −10.208 |
| | $^{145}$Nd($p$,$\alpha 3n$) | −15.964 |
| | $^{146}$Nd($p$,$\alpha 4n$) | −23.529 |
| $^{140}$Pr | Decay of $^{140}$Nd | |
| | $^{142}$Nd($p$,$^{3}$He) | −8.902 |
| | $^{143}$Nd($p$,n$^{3}$He) | −15.026 |
| | $^{144}$Nd($p$,2$n^{3}$He) | −22.843 |
| $^{142}$Pr+$^{142}$Pr$^{m}$ | $^{143}$Nd($p$,2$p$) | −7.504 |
| | $^{144}$Nd($p$,$^{3}$He) | −7.603 |
| | $^{145}$Nd($p$,$\alpha$) | +7.220 |
| | $^{146}$Nd($p$,$\alpha n$) | −0.346 |
| $^{147}$Pr | $^{148}$Nd($p$,2$p$) | −9.248 |
| | $^{150}$Nd($p$,$\alpha$) | +6.629 |
| $^{149}$Pr | $^{150}$Nd($p$,2$p$) | −9.922 |
| $^{139}$Ce | Decay of $^{139}$Pr | |

TABLE III. Half-lives, decay modes, main gamma lines and their intensities of the direct and indirect activation products of the $^{nat}Nd(p,x)$ reactions and of the monitoring reactions.

| RN | Half-life | Decay mode | $E_\gamma$ in keV ($I_\gamma$ in %) |
|---|---|---|---|
| $^{140}Pm$ | 9.2 s | EC (100 %) | 773.8 (5.0) |
| $^{140}Pm^m$ | 5.95 min | EC (100 %) | 419.57 (92.0), 773.74 (100), 1028.19 (100) |
| $^{141}Pm$ | 20.90 min | EC (100 %) | 193.68 (1.61), 622.01 (0.85), 886.22 (2.44) 1223.26 (4.74), 1345.52 (1.33) |
| $^{143}Pm$ | 265 d | EC (100 %) | 741.98 (38.5) |
| $^{144}Pm$ | 363 d | EC (100 %) | 476.80 (43.8), 618.01 (98), 696.51 (99.49) |
| $^{146}Pm$ | 5.53 a | EC (66.0 %) $\beta^-$ (34.0 %) | 453.88 (65), 735.72 (22.5), 747.16 (34) |
| $^{148}Pm$ | 5.368 d | $\beta^-$ (100 %) | 550.28 (22), 914.85 (11.46), 1465.12 (22.2) |
| $^{148}Pm^m$ | 41.29 d | IT (4.2 %) $\beta^-$ (95.8 %) | 288.14 (12.56), 414.03 (18.66), 550.28 (94.9) 599.81 (12.54), 629.99 (89), 725.67 (32.8) 915.33 (17.17), 1013.81 (20.3) |
| $^{149}Pm$ | 53.08 hr | $\beta^-$ (100 %) | 285.95 (3.1) |
| $^{150}Pm$ | 2.68 hr | $\beta^-$ (100 %) | 333.97 (68), 406.52 (5.6), 831.92 (11.9), 876.41 (7.3), 1165.74 (15.8), 1324.51 (17.5) |
| $^{140}Nd$ | 3.37 d | EC (100 %) | no gammas |
| $^{141}Nd$ | 2.49 hr | EC (100 %) | 145.44 (0.239), 1126.80 (0.8), 1147.20 (0.306), 1292.60 (0.46), 1298.60 (0.127) |
| $^{147}Nd$ | 10.98 d | $\beta^-$ (100 %) | 91.11 (28.1), 531.02 (13.4) |
| $^{149}Nd$ | 1.728 hr | $\beta^-$ (100 %) | 114.31 (19.2), 211.31 (25.9), 267.69 (6.03), 270.17 (10.7), 326.55 (4.56), 423.55 (7.4) 540.51 (6.58), 654.83 (8) |
| $^{138}Pr^m$ | 2.12 hr | EC (100 %) | 302.70 (80), 788.74 (100), 1037.80 (101) |
| $^{139}Pr$ | 4.41 hr | EC (100 %) | 1347.33 (0.473), 1375.56 (0.154), 1630.67 (0.343) |
| $^{140}Pr$ | 3.39 min | EC (100 %) | 306.9 (0.147), 1596.1 (0.49) |
| $^{142}Pr$ | 19.12 hr | $\beta^-$ (99.9836 %) EC (0.0164 %) | 1575.6 (3.7) |
| $^{147}Pr$ | 13.4 min | $\beta^-$ (100 %) | 314.7 (17.5), 577.9 (13.7), 641.4 (15.6) |
| $^{149}Pr$ | 2.26 min | $\beta^-$ (100 %) | 108.52 (9.5), 138.45 (11.0), 165.09 (9.9), 258.33 (5.7), 332.94 (6.15), 517.44 (4.80) |
| $^{139}Ce$ | 137.641 d | EC (100 %) | 165.86 (80) |
| $^{48}V$ | 15.9735 d | EC (100 %) | 983.50 (0.9998), 1312.10 (0.975) |
| $^{62}Zn$ | 9.187 hr | EC (100 %) | 548.35 (15.2), 596.56 (25.7) |
| $^{63}Zn$ | 38.47 min | EC (100 %) | 669.62 (8.4), 962.06 (6.6) |
| $^{65}Zn$ | 244.1 d | EC (100 %) | 1115.55 (50.75) |

TABLE IV. Measured cross sections for formation of $^{141}$Pm, $^{143}$Pm, $^{144}$Pm, $^{146}$Pm, $^{148}$Pm, $^{148m}$Pm, $^{149}$Pm, $^{149}$Pm$^{cum}$ and $^{150}$Pm via $^{nat}$Nd($p$,$x$) reactions.

| $E_p$ (MeV) | Cross Section (mb) | | | | | | | | |
|---|---|---|---|---|---|---|---|---|---|
| | $^{141}$Pm | $^{143}$Pm | $^{144}$Pm | $^{146}$Pm | $^{148}$Pm | $^{148}$Pm$^{m}$ | $^{149}$Pm | $^{149}$Pm$^{cum}$ | $^{150}$Pm |
| 29.04 | 183 ± 20 | 144 ± 16 | 112 ± 12 | 22.5 ± 2.5 | 4.86 ± 0.53 | 10.7 ± 1.2 | 3.64 ± 0.43 | 9.76 ± 1.07 | 0.814 ± 0.090 |
| 27.66 | 215 ± 23 | 135 ± 15 | 140 ± 15 | 28.6 ± 3.1 | 6.56 ± 0.72 | 14.7 ± 1.6 | 3.91 ± 0.46 | 10.1 ± 1.1 | 0.869 ± 0.096 |
| 26.22 | 212 ± 23 | 131 ± 14 | 156 ± 17 | 37.0 ± 4.1 | 9.11 ± 1.00 | 19.5 ± 2.1 | 4.23 ± 0.49 | 10.1 ± 1.1 | 0.919 ± 0.101 |
| 25.18 | 226 ± 25 | 167 ± 18 | 182 ± 20 | 47.4 ± 5.2 | 13.4 ± 1.5 | 26.8 ± 2.9 | 5.64 ± 0.66 | 12.0 ± 1.3 | 1.09 ± 0.12 |
| 23.63 | 204 ± 22 | 216 ± 24 | 177 ± 19 | 50.8 ± 5.6 | 18.5 ± 2.0 | 31.5 ± 3.4 | 6.82 ± 0.80 | 12.9 ± 1.4 | 1.14 ± 0.13 |
| 22.00 | 181 ± 20 | 253 ± 28 | 155 ± 17 | 46.2 ± 5.0 | 20.5 ± 2.2 | 30.0 ± 3.3 | 8.93 ± 1.04 | 14.1 ± 1.5 | 1.22 ± 0.13 |
| 21.38 | 193 ± 21 | 268 ± 29 | 161 ± 18 | 48.7 ± 5.4 | 21.7 ± 2.4 | 31.1 ± 3.4 | 9.94 ± 1.16 | 15.3 ± 1.7 | 1.29 ± 0.14 |
| 20.10 | 173 ± 19 | 253 ± 28 | 130 ± 14 | 40.0 ± 4.4 | 20.2 ± 2.2 | 26.0 ± 2.8 | 13.8 ± 1.6 | 18.4 ± 2.0 | 1.34 ± 0.15 |
| 19.08 | 156 ± 17 | 238 ± 26 | 104 ± 11 | 30.9 ± 3.4 | 18.4 ± 2.0 | 20.8 ± 2.3 | 19.9 ± 2.3 | 23.6 ± 2.6 | 1.41 ± 0.15 |
| 18.60 | 171 ± 19 | 256 ± 28 | 113 ± 12 | 34.1 ± 3.8 | 19.8 ± 2.2 | 22.8 ± 2.5 | 21.7 ± 2.5 | 25.8 ± 2.8 | 1.51 ± 0.17 |
| 17.52 | 143 ± 16 | 237 ± 26 | 95.9 ± 10.5 | 21.0 ± 2.4 | 15.4 ± 1.7 | 15.3 ± 1.7 | 32.5 ± 3.8 | 35.7 ± 3.9 | 1.63 ± 0.18 |
| 16.38 | 83.2 ± 9.1 | 218 ± 24 | 93.7 ± 10.2 | 8.95 ± 1.05 | 7.82 ± 0.86 | 6.44 ± 0.70 | 44.6 ± 5.2 | 46.8 ± 5.1 | 1.66 ± 0.18 |
| 15.50 | 53.6 ± 5.9 | 206 ± 23 | 91.0 ± 9.9 | 6.32 ± 0.79 | 4.67 ± 0.51 | 3.51 ± 0.38 | 47.8 ± 5.6 | 49.5 ± 5.4 | 1.68 ± 0.18 |
| 14.25 | 1.43 ± 0.20 | 200 ± 22 | 86.5 ± 9.4 | 7.36 ± 0.86 | 1.11 ± 0.12 | 0.886 ± 0.098 | 48.8 ± 5.7 | 49.7 ± 5.4 | 1.79 ± 0.19 |
| 12.92 | | 183 ± 20 | 85.1 ± 9.3 | 10.2 ± 1.2 | 1.16 ± 0.13 | 0.928 ± 0.103 | 39.7 ± 4.6 | 39.9 ± 4.4 | 1.92 ± 0.21 |
| 11.53 | | 142 ± 15 | 100 ± 11 | 16.6 ± 1.8 | 1.73 ± 0.19 | 1.29 ± 0.14 | 34.1 ± 4.0 | 34.1 ± 3.7 | 2.43 ± 0.26 |
| 9.96 | | 43.2 ± 4.7 | 94.2 ± 10.3 | 34.8 ± 3.8 | 3.48 ± 0.38 | 1.91 ± 0.21 | 19.0 ± 2.2 | 19.0 ± 2.1 | 3.54 ± 0.39 |

TABLE V. Measured cross sections for formation of $^{140}Nd^{cum}$, $^{141}Nd^{cum}$, $^{147}Nd^{cum}$, $^{149}Nd$, $^{138}Pr^{m}$, $^{139}Pr$, $^{142}Pr^{cum}$, and $^{139}Ce^{cum}$ via $^{nat}Nd(p,x)$ reactions

| $E_p$ (MeV) | Cross section (mb) | | | | | | | |
|---|---|---|---|---|---|---|---|---|
| | $^{140}Nd^{cum}$ | $^{141}Nd^{cum}$ | $^{147}Nd^{cum}$ | $^{149}Nd$ | $^{138}Pr^{m}$ | $^{139}Pr$ | $^{142}Pr^{cum}$ | $^{139}Ce^{cum}$ |
| 29.04 | 75.5 ± 8.3 | 128 ± 15 | 6.44 ± 0.71 | 5.92 ± 0.65 | 1.00 ± 0.11 | 3.07 ± 0.34 | 0.986 ± 0.131 | 2.97 ± 0.32 |
| 27.66 | 29.0 ± 3.2 | 130 ± 15 | 6.44 ± 0.70 | 5.98 ± 0.65 | 0.736 ± 0.082 | 2.71 ± 0.30 | 0.955 ± 0.115 | 2.61 ± 0.28 |
| 26.22 | 4.67 ± 0.56 | 123 ± 15 | 6.09 ± 0.66 | 5.67 ± 0.62 | 0.475 ± 0.053 | 2.28 ± 0.25 | 0.948 ± 0.110 | 2.19 ± 0.24 |
| 25.18 | 1.94 ± 0.34 | 132 ± 16 | 6.56 ± 0.72 | 6.18 ± 0.67 | 0.380 ± 0.042 | 2.34 ± 0.26 | 1.10 ± 0.13 | 2.25 ± 0.25 |
| 23.63 | 1.20 ± 0.23 | 118 ± 14 | 6.11 ± 0.67 | 5.86 ± 0.64 | 0.186 ± 0.023 | 2.14 ± 0.24 | 0.995 ± 0.117 | 2.06 ± 0.23 |
| 22.00 | | 93.0 ± 11.0 | 5.19 ± 0.57 | 5.00 ± 0.55 | 0.0564 ± 0.0079 | 1.85 ± 0.20 | 0.794 ± 0.096 | 1.78 ± 0.20 |
| 21.38 | | 111 ± 13 | 5.31 ± 0.58 | 5.19 ± 0.57 | 0.0663 ± 0.0094 | 1.88 ± 0.21 | 0.819 ± 0.110 | 1.88 ± 0.21 |
| 20.10 | | 89.0 ± 10.6 | 4.30 ± 0.48 | 4.39 ± 0.48 | | 1.53 ± 0.17 | 0.643 ± 0.092 | 1.53 ± 0.17 |
| 19.08 | | 74.7 ± 8.9 | 3.81 ± 0.42 | 3.64 ± 0.40 | | 1.16 ± 0.13 | 0.525 ± 0.073 | 1.17 ± 0.13 |
| 18.60 | | 84.5 ± 10.1 | 4.02 ± 0.44 | 3.95 ± 0.43 | | 1.27 ± 0.14 | | 1.28 ± 0.14 |
| 17.52 | | 70.5 ± 8.4 | 3.08 ± 0.34 | 3.06 ± 0.33 | | 0.940 ± 0.105 | | 0.942 ± 0.106 |
| 16.38 | | 35.9 ± 4.2 | 2.02 ± 0.23 | 2.09 ± 0.23 | | 0.563 ± 0.065 | | 0.564 ± 0.065 |
| 15.50 | | 19.9 ± 2.3 | 1.56 ± 0.18 | 1.66 ± 0.18 | | 0.475 ± 0.054 | | 0.475 ± 0.054 |
| 14.25 | | 1.46 ± 0.18 | 0.703 ± 0.085 | 0.794 ± 0.086 | | 0.231 ± 0.029 | | 0.232 ± 0.029 |
| 12.92 | | | 0.272 ± 0.046 | 0.250 ± 0.027 | | 0.103 ± 0.017 | | 0.103 ± 0.017 |
| 11.53 | | | | 0.0584 ± 0.0071 | | 0.0474 ± 0.0125 | | 0.0458 ± 0.0121 |
| 9.96 | | | | 0.00493 ±0.0014 | | | | |

Table VI. Estimated production rates (in atoms per kg of Nd and day) and activities of the long-lived radionuclides due to cosmogenic proton activation in the energy range of 10–30 MeV. Both experiment-based and TENDL-based calculations are displayed. The activities were calculated for an exposure time of 30 days, a cooling time underground of 30 days and 1000 kg of natural neodymium (0.1 % of $^{nat}Nd$ loaded in 1 kton of scintillator). The total proton flux at the sea level per energy unit was adopted from [8].

| Radionuclide | Half-life (d) | Production rate ($kg^{-1}d^{-1}$) , TENDL | Production rate ($kg^{-1}d^{-1}$) , Expt. | A (mBq) , TENDL 10–30 MeV | A (mBq) , Expt. 10–30 MeV | $A_{TENDL}/A_{Expt}$ 10–30 MeV |
|---|---|---|---|---|---|---|
| $^{143}Pm$ | 265 | 0.1665 | 0.1260 | 0.1345 | 0.1018 | 1.32 |
| $^{144}Pm$ | 363 | 0.0967 | 0.0830 | 0.0588 | 0.0505 | 1.16 |
| $^{146}Pm$ | 2020 | 0.0255 | 0.0196 | 0.0030 | 0.0023 | 1.30 |
| $^{148}Pm$ | 5.37 | 0.0134 | 0.0073 | 0.0032 | 0.0017 | 1.82 |
| $^{148}Pm^{m}$ | 41.29 | 0.0067 | 0.0109 | 0.0185 | 0.0301 | 0.62 |
| $^{147}Nd$ | 10.98 | 0.0027 | 0.0026 | 0.0039 | 0.0038 | 1.04 |
| $^{139}Pr$ | 0.184 | 0.0017 | 0.0001 | 0.0000 | 0.0000 | 13.66 |
| $^{139}Ce$ | 137.64 | decay of $^{139}Pr$ | decay of $^{139}Pr$ | 0.0023 | 0.0002 | 13.66 |

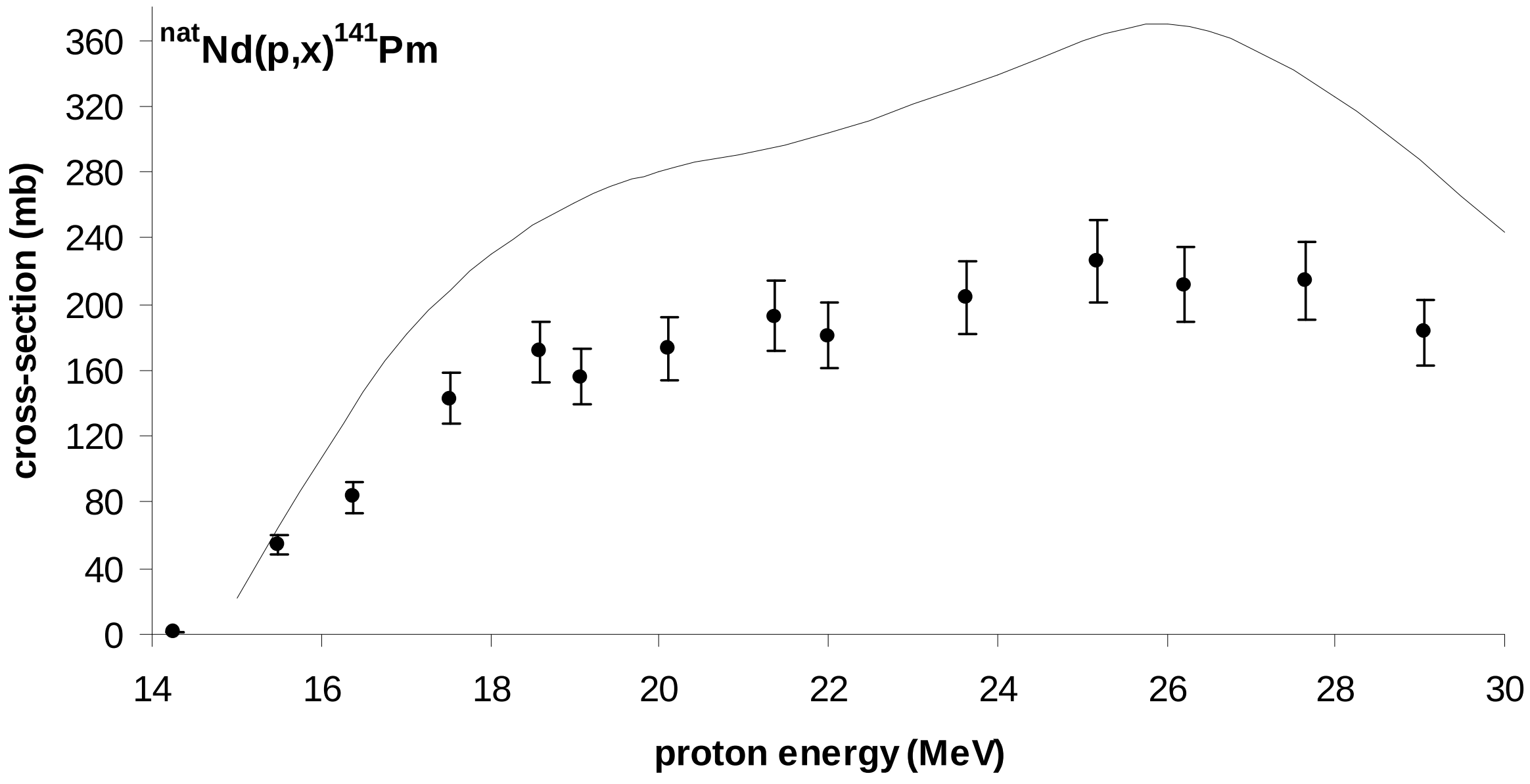

FIG. 1. Experimental cross sections for the $^{nat}Nd(p,x)^{141}Pm$ reactions compared to the TENDL database.

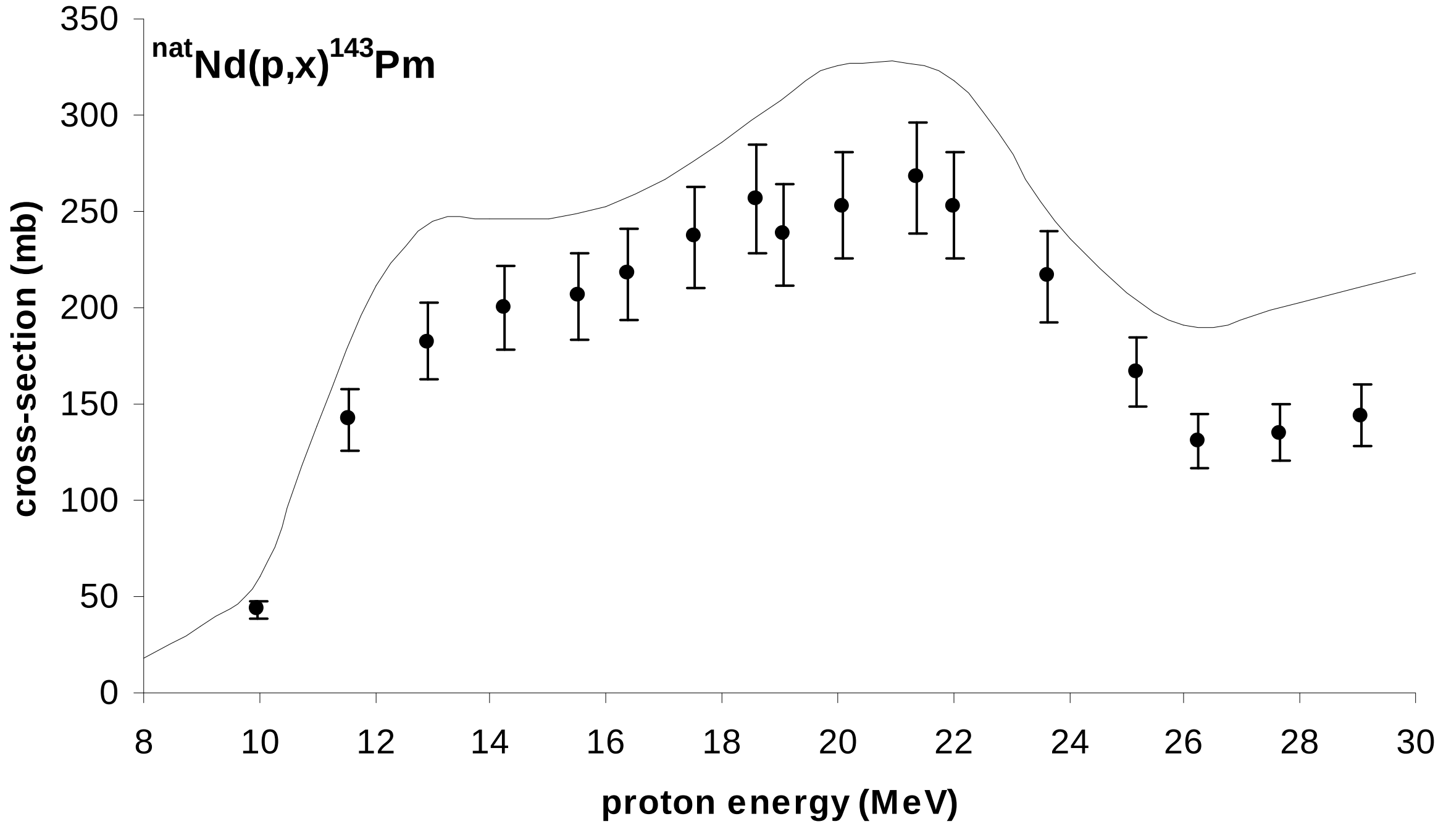

FIG. 2. Experimental cross sections for the $^{nat}Nd(p,x)^{143}Pm$ reactions compared to the TENDL database.

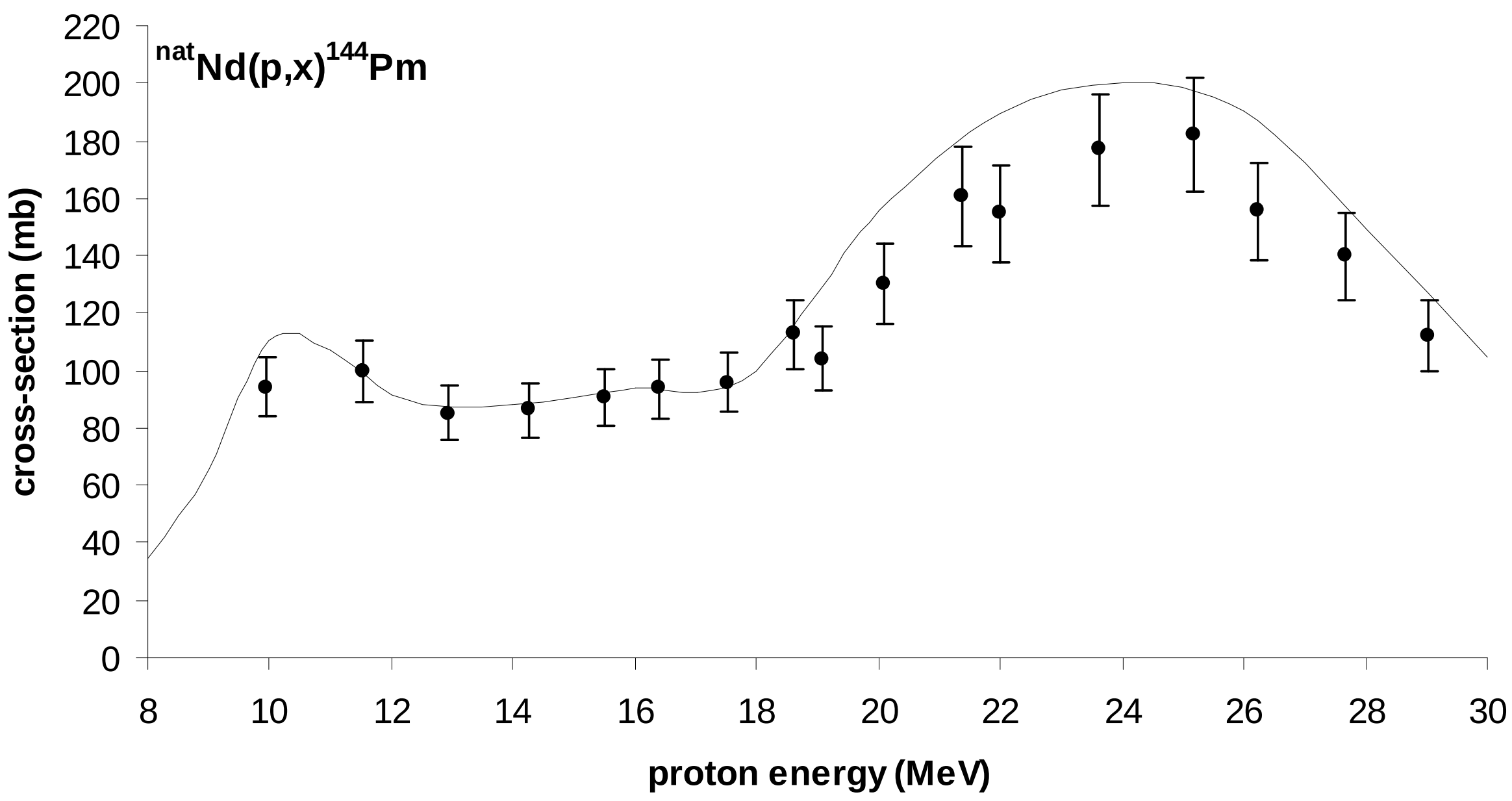

FIG. 3. Experimental cross sections for the $^{nat}Nd(p,x)^{144}Pm$ reactions compared to the TENDL database

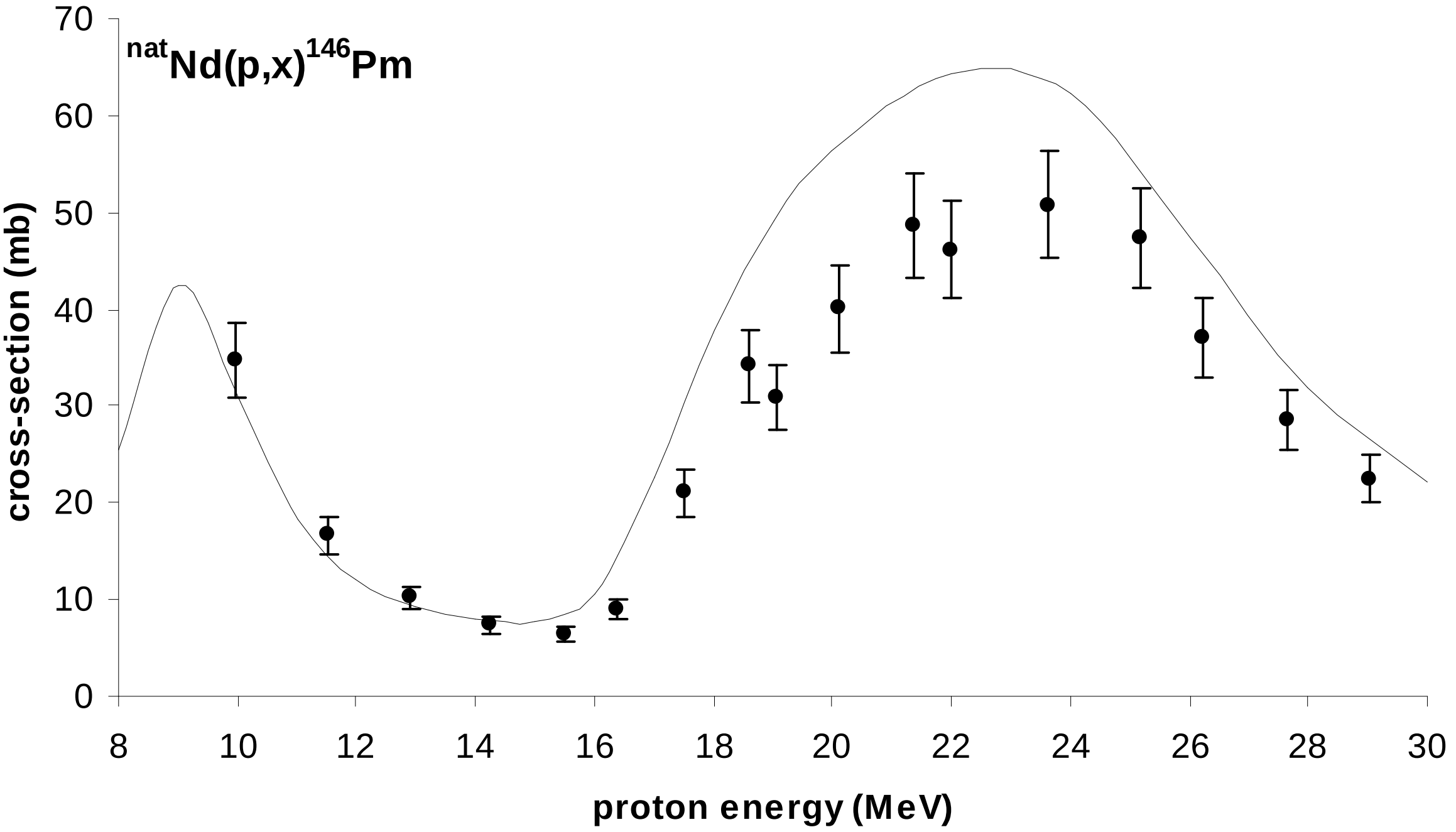

FIG. 4. Experimental cross sections for the $^{nat}Nd(p,x)^{146}Pm$ reactions compared to the TENDL database.

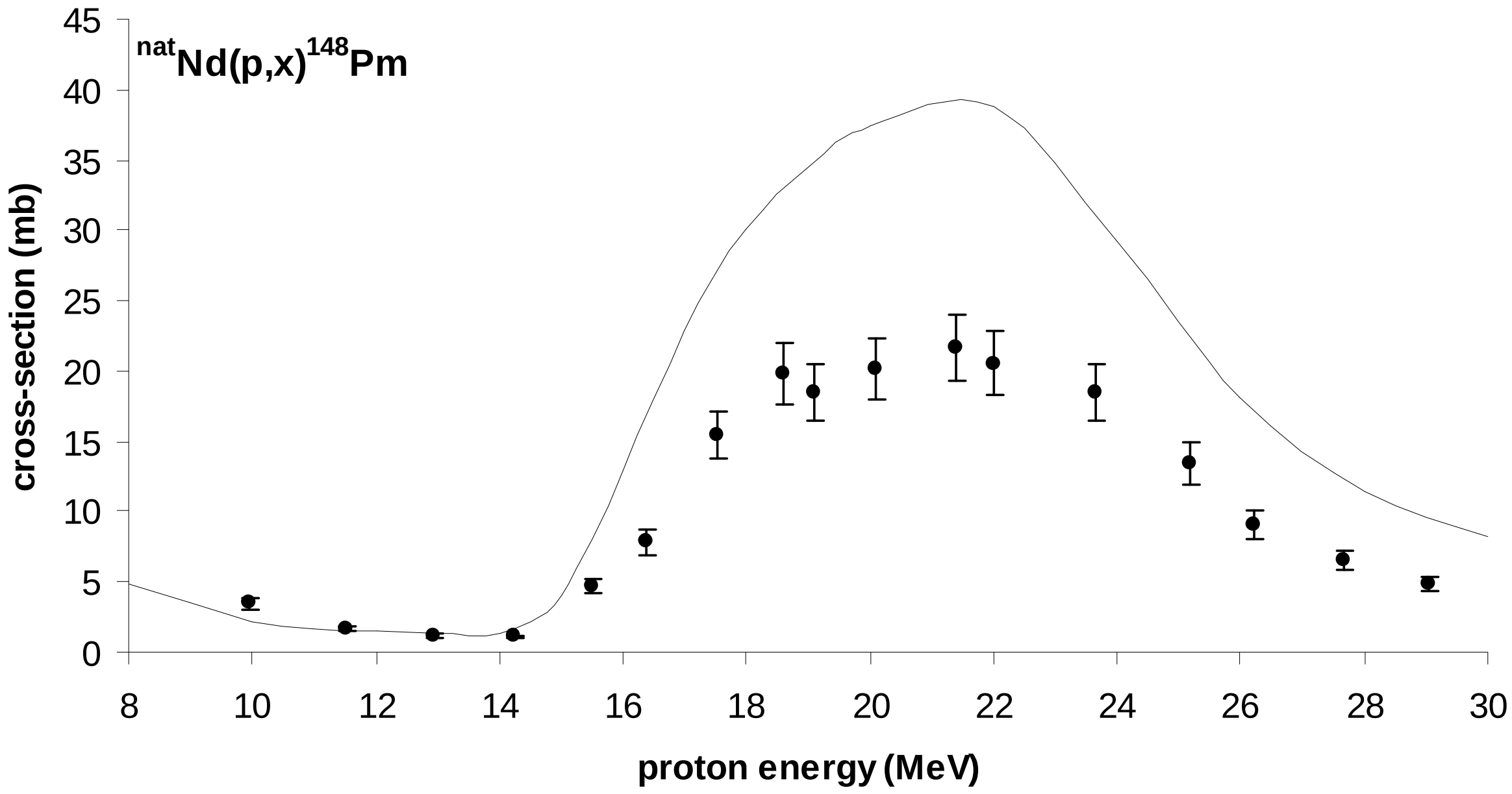

FIG. 5. Experimental cross sections for the $^{nat}Nd(p,x)^{148}Pm$ reactions compared to the TENDL database.

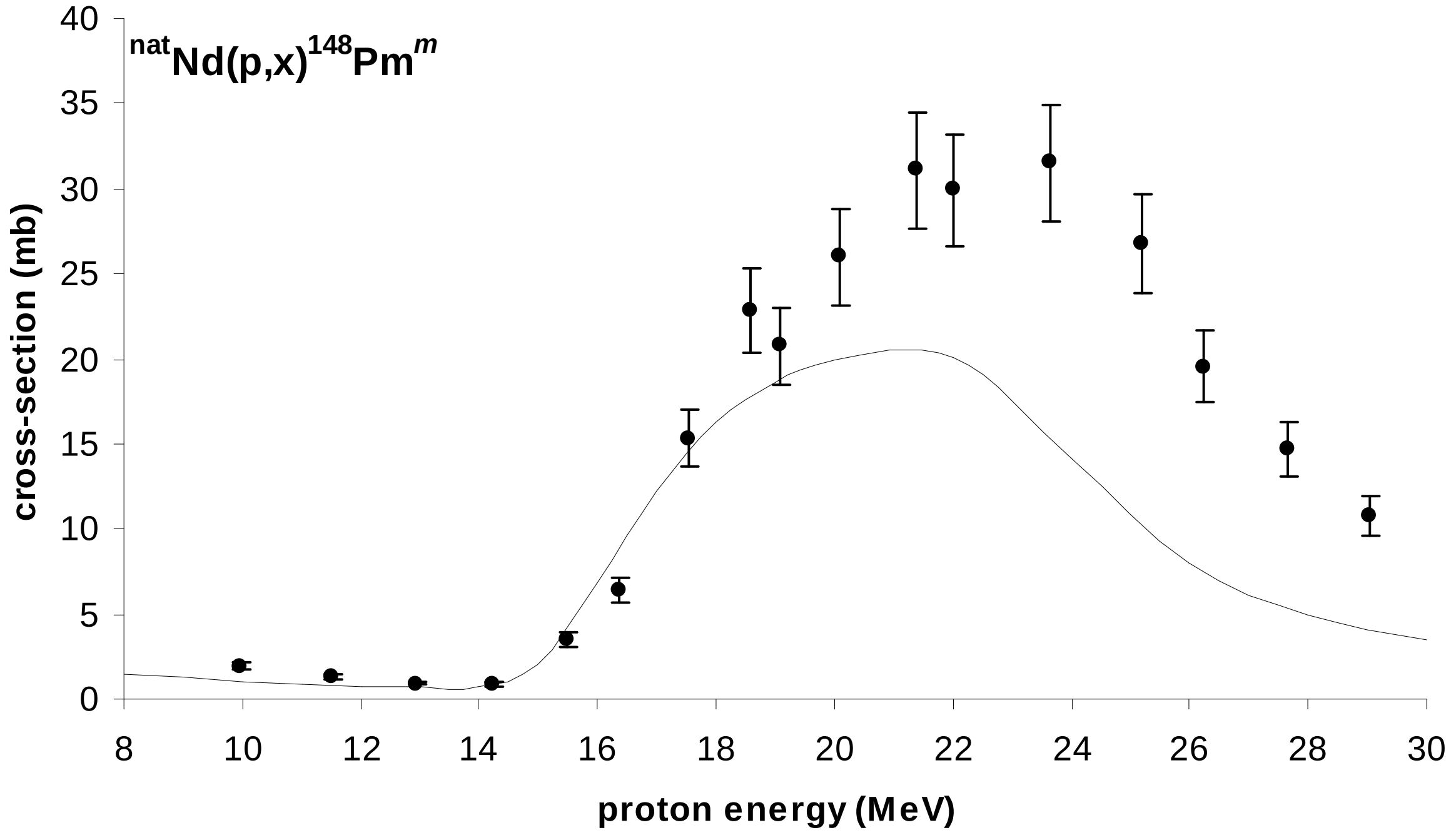

FIG. 6. Experimental cross sections for the $^{nat}Nd(p,x)^{148}Pm^{m}$ reactions compared to the TENDL database.

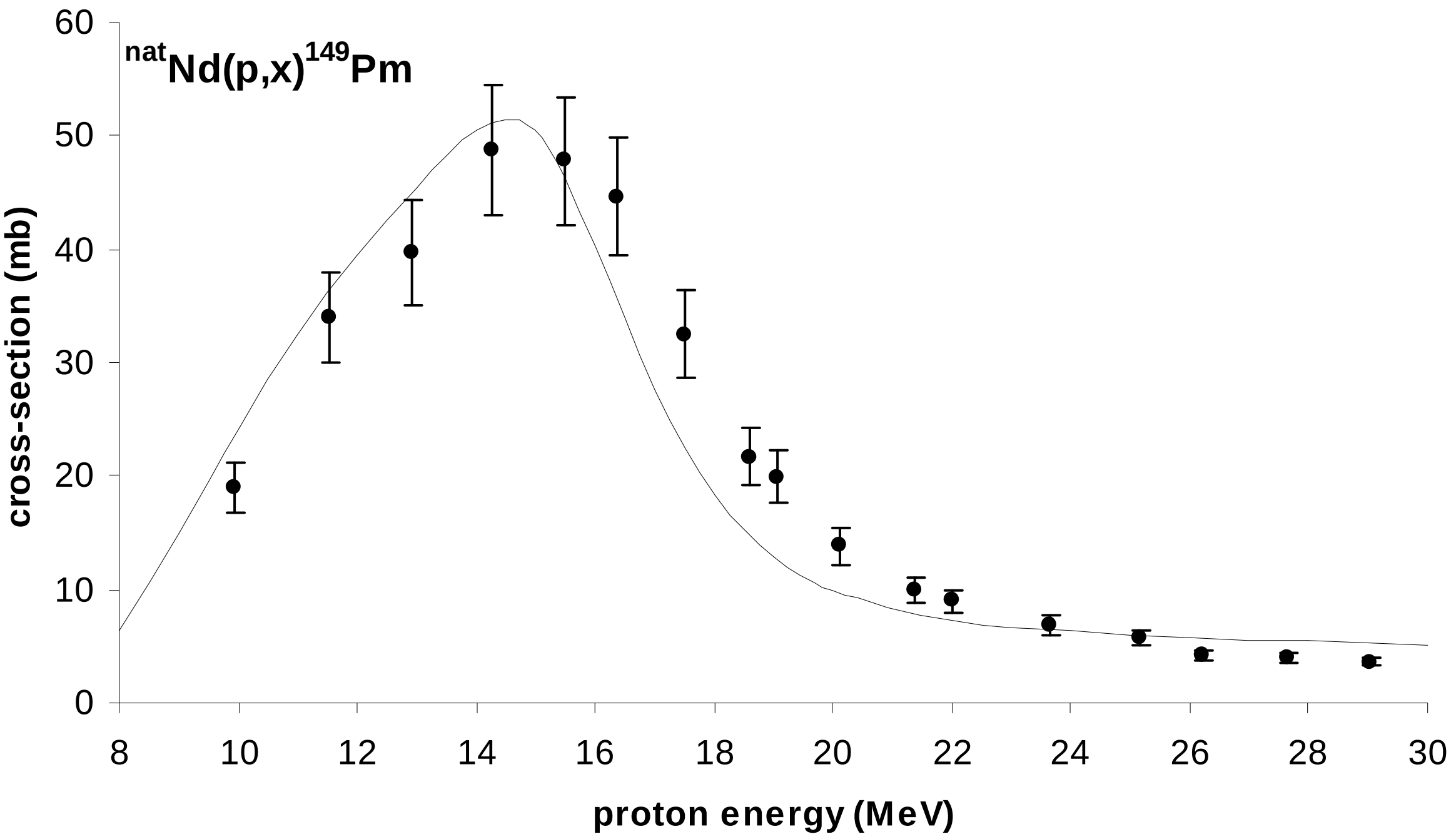

FIG. 7. Experimental cross sections for the ${}^{nat}$Nd($p$,$x$)${}^{149}$Pm reactions compared to the TENDL database.

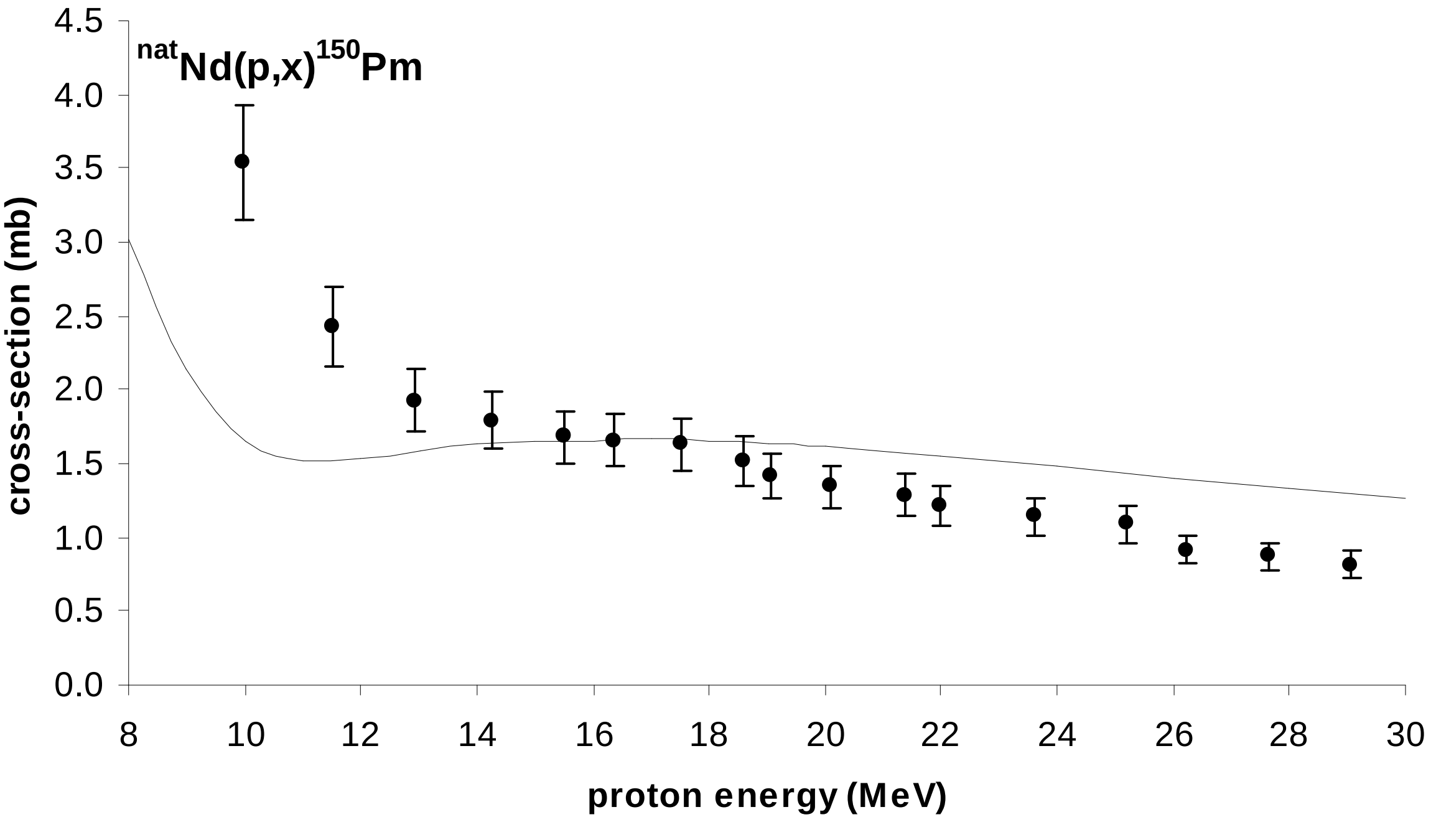

FIG. 8. Experimental cross sections for the ${}^{nat}$Nd($p$,$x$)${}^{150}$Pm reactions compared to the TENDL database.

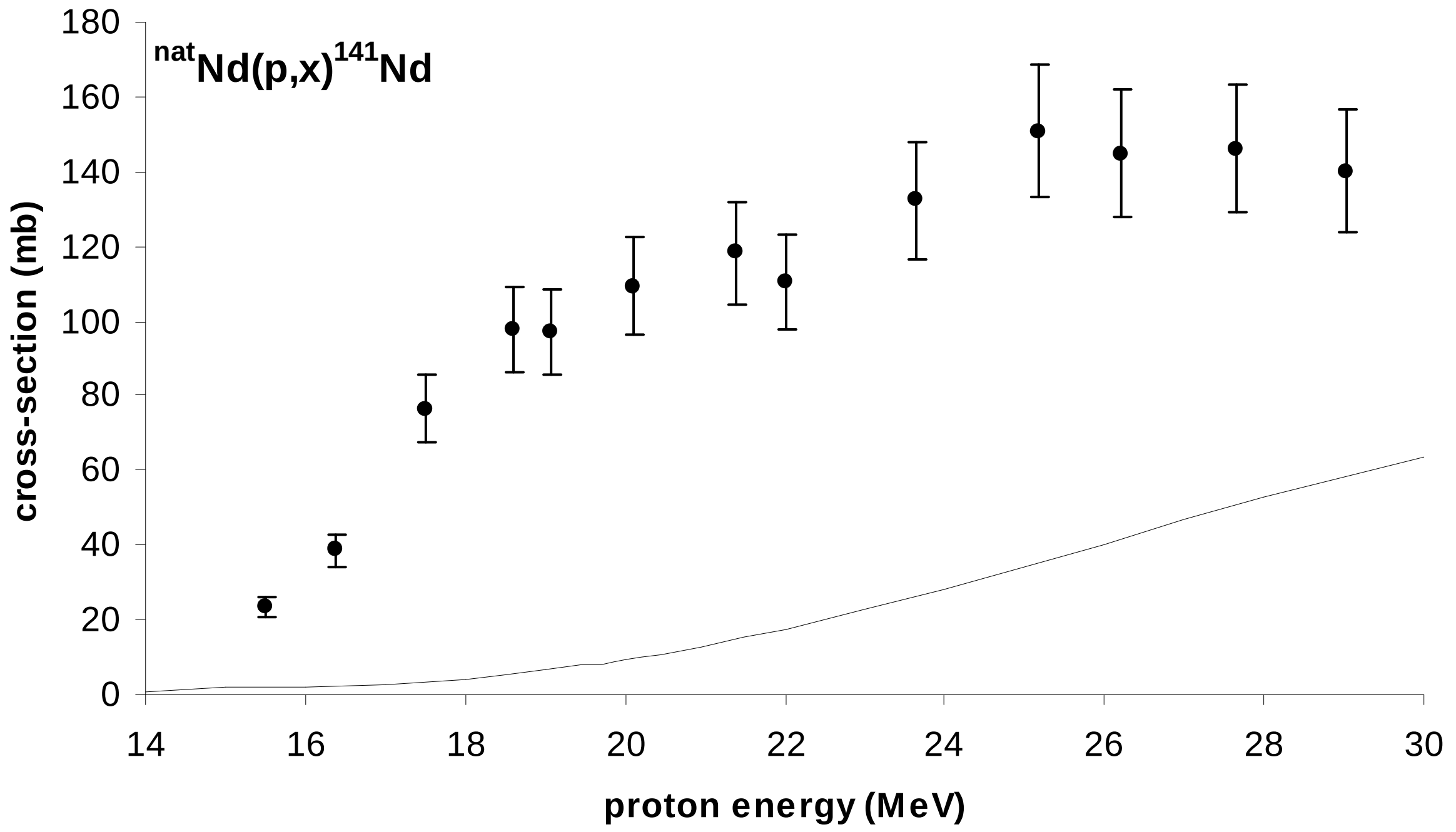

FIG. 9. Experimental cross sections for the ${}^{nat}$Nd($p$,$x$)${}^{141}$Nd reactions compared to the TENDL database.

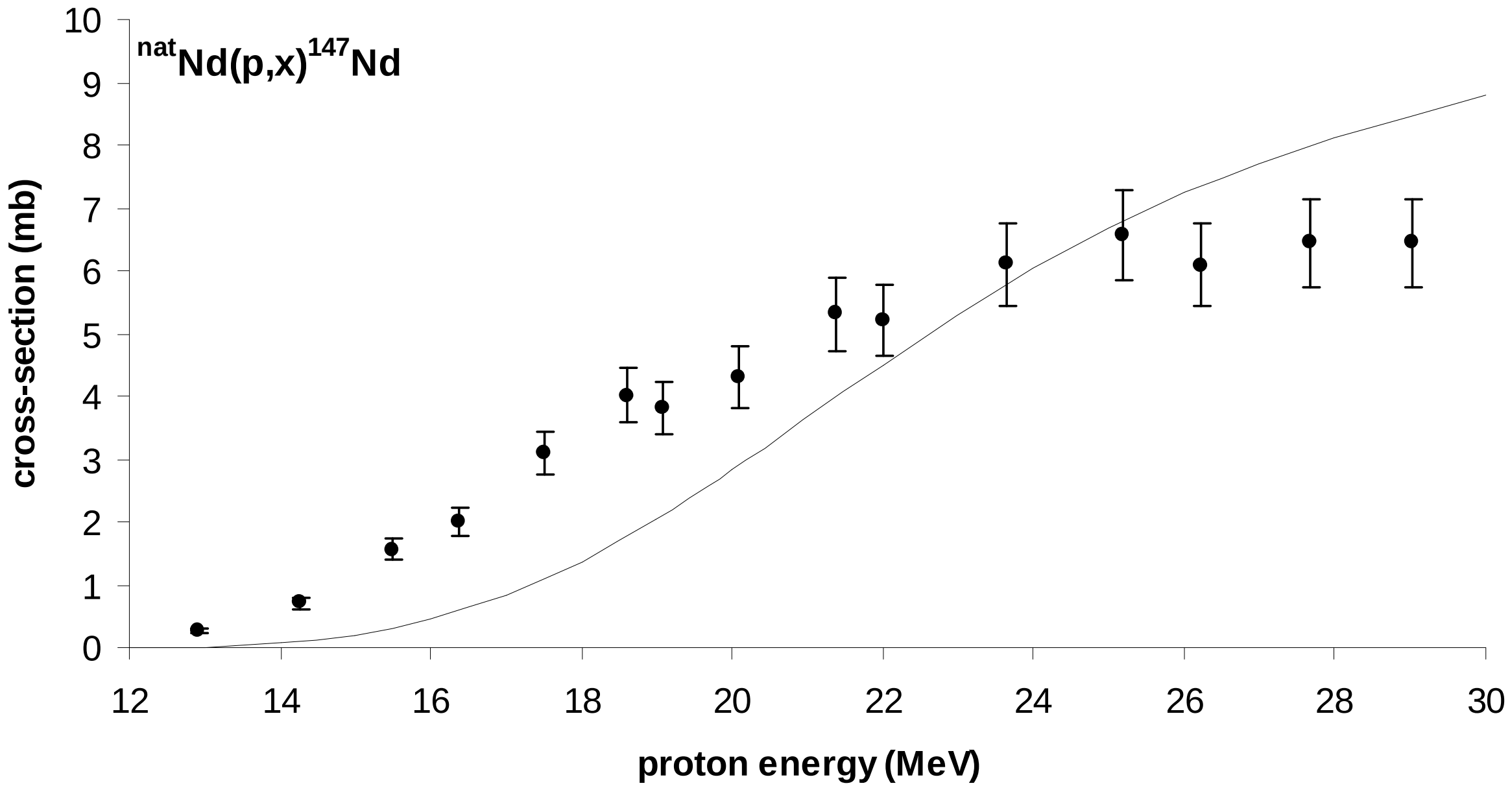

FIG. 10. Experimental cross sections for the ${}^{nat}$Nd($p$,$x$)${}^{147}$Nd reactions compared to the TENDL database.

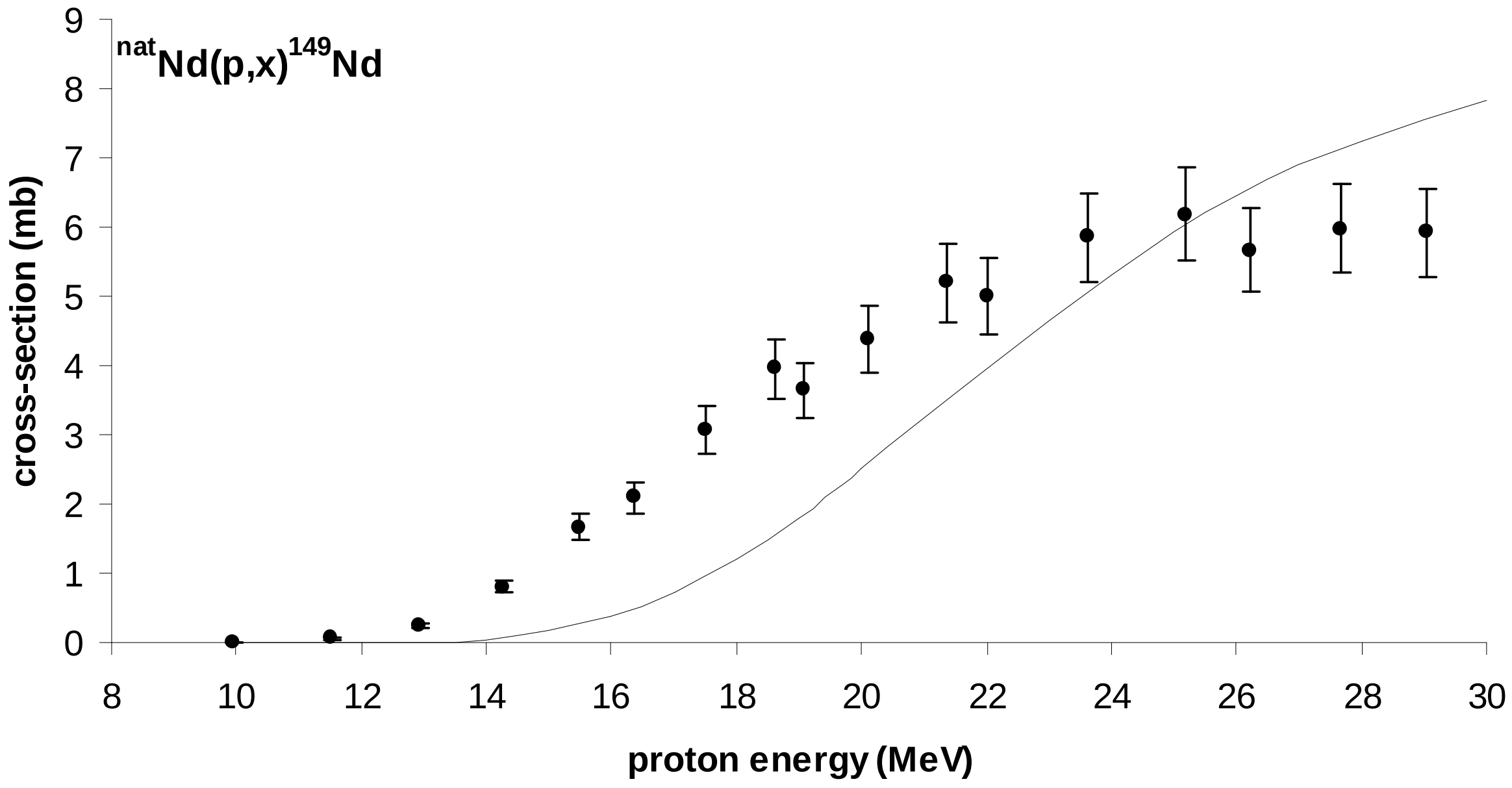

FIG. 11. Experimental cross sections for the $^{nat}$Nd($p$,$x$)$^{149}$Nd reactions compared to the TENDL database.

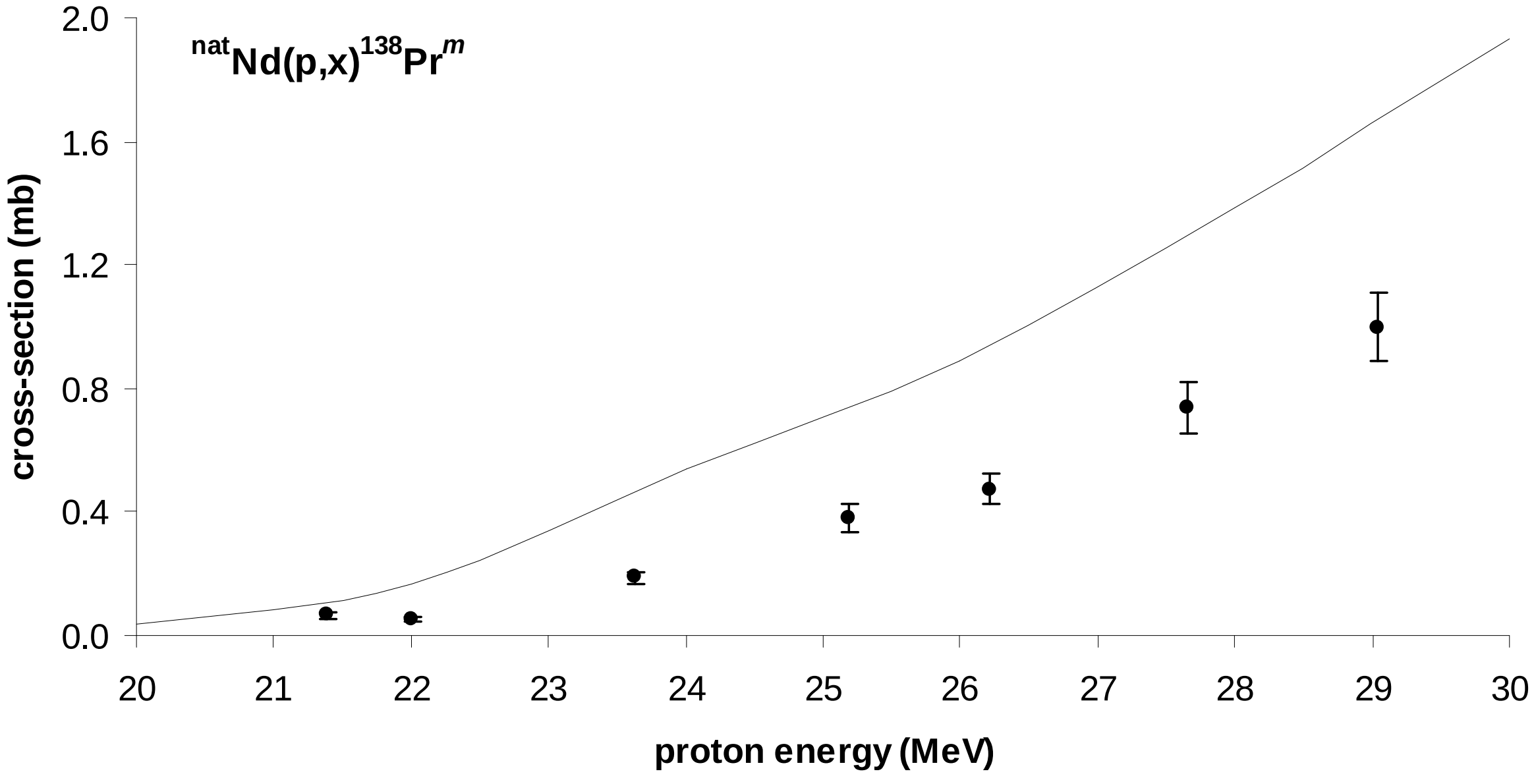

FIG. 12. Experimental cross sections for the $^{nat}$Nd($p$,$x$)$^{138}$Pr$^{m}$ reactions compared to the TENDL database.

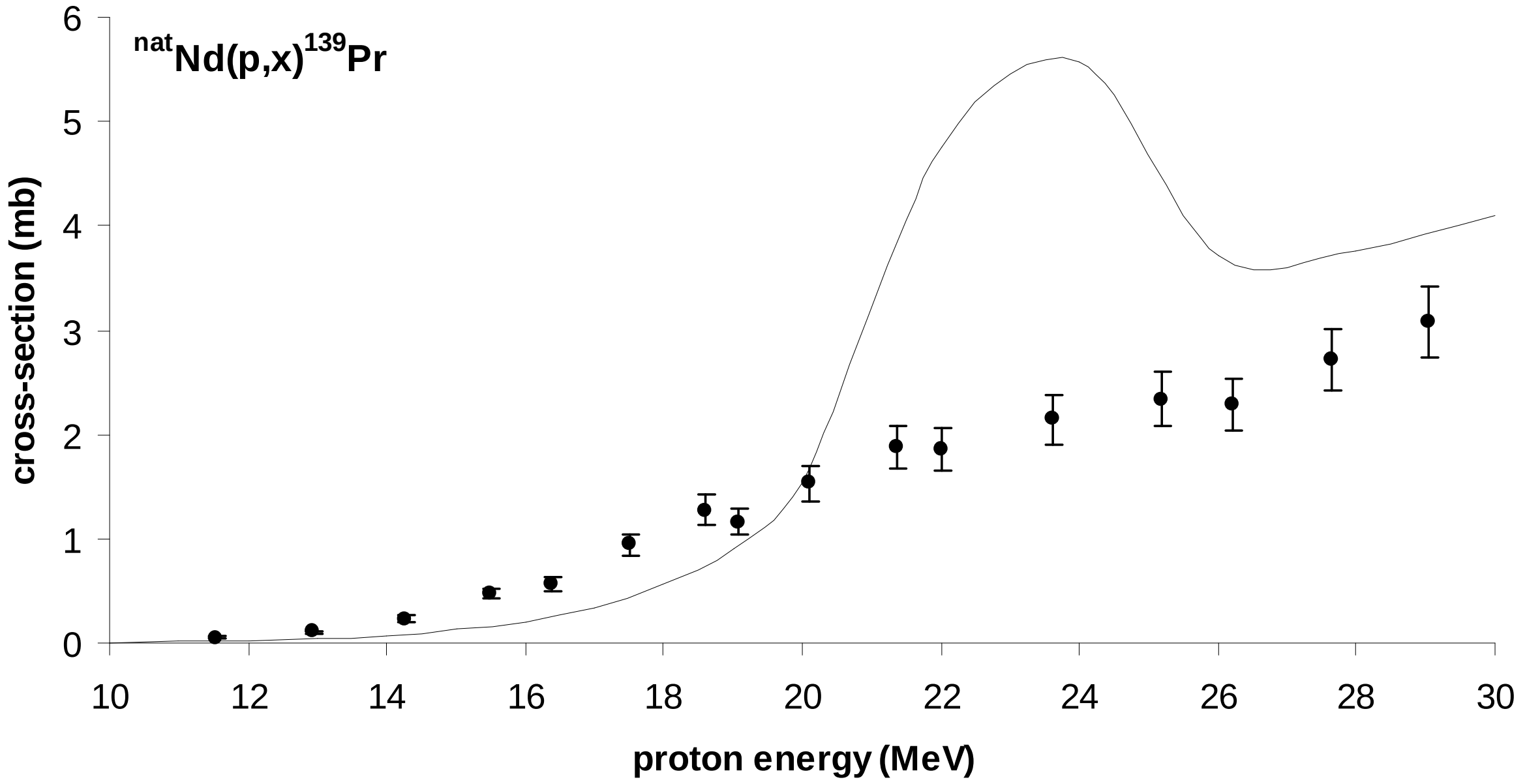

FIG. 13. Experimental cross sections for the ${}^{nat}$Nd(*p*,*x*)${}^{139}$Pr reactions compared to the TENDL database.

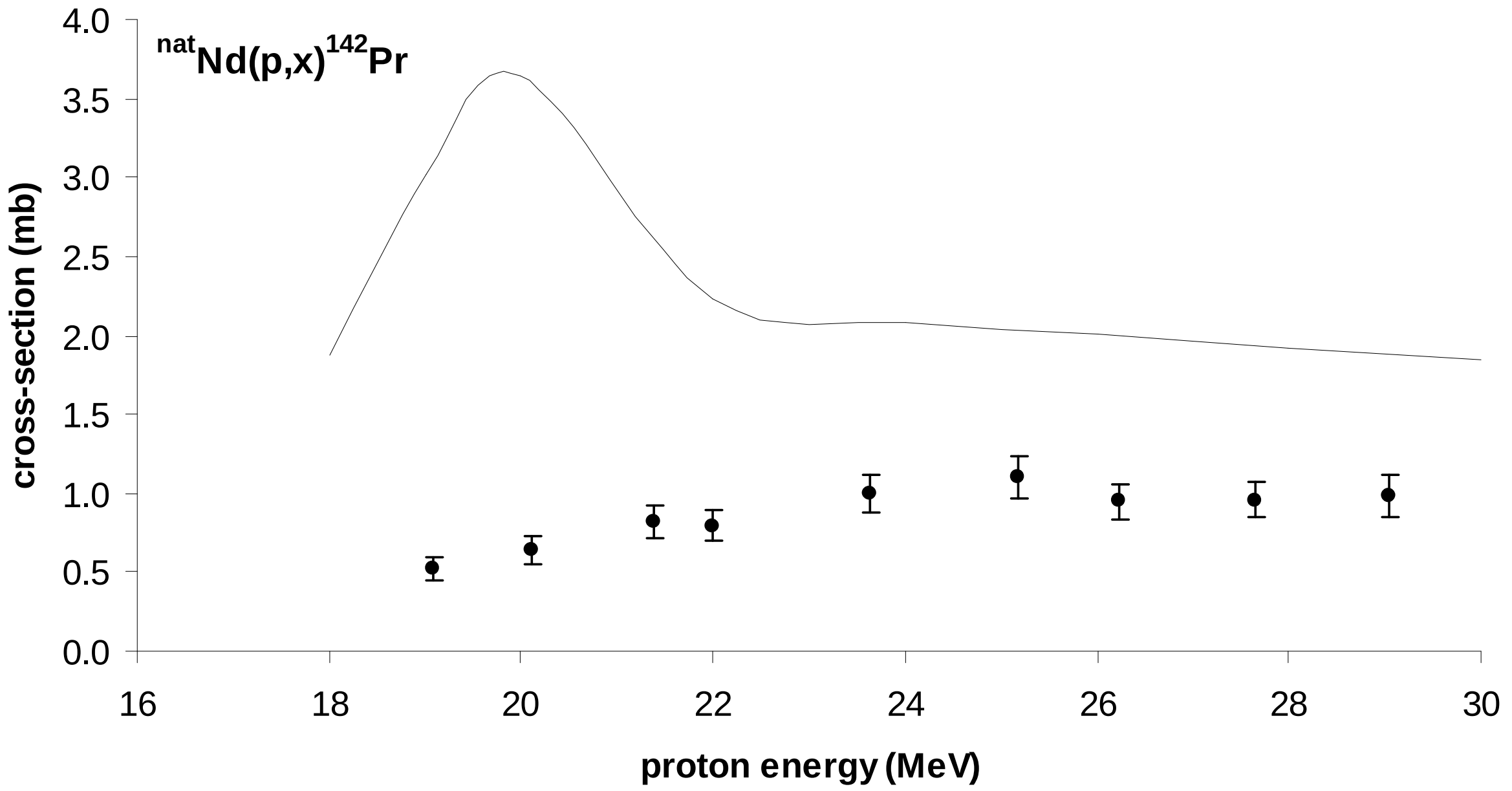

FIG. 14. Experimental cross sections for the ${}^{nat}$Nd(*p*,*x*)${}^{142}$Pr reactions compared to the TENDL database.

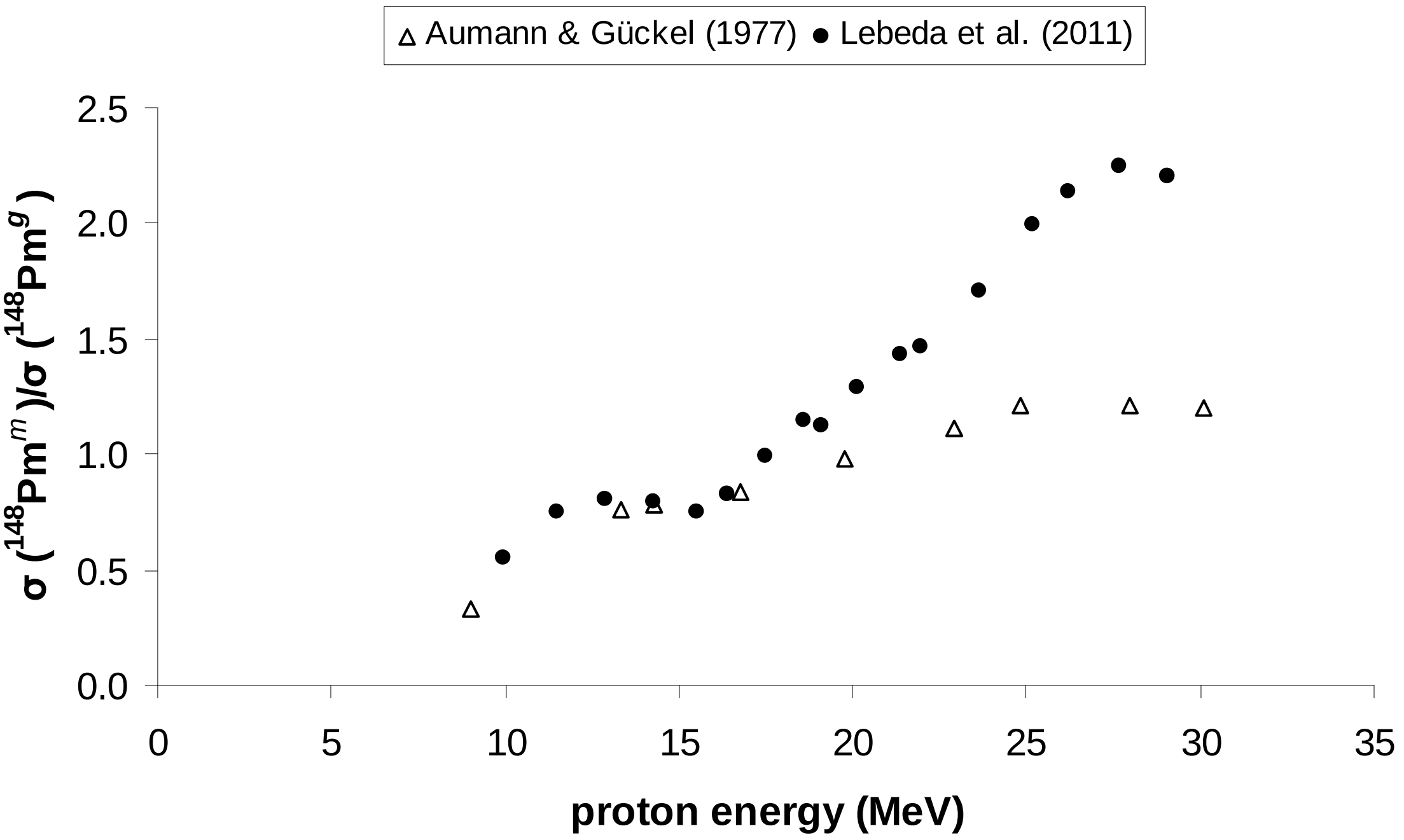

FIG. 15. Experimental isomeric cross-section ratios for the $^{148}$Nd($p$,$n$) reaction measured by Aumann and Gückel (1977) and for the $^{nat}$Nd($p$,$x$) reactions measured in this work.

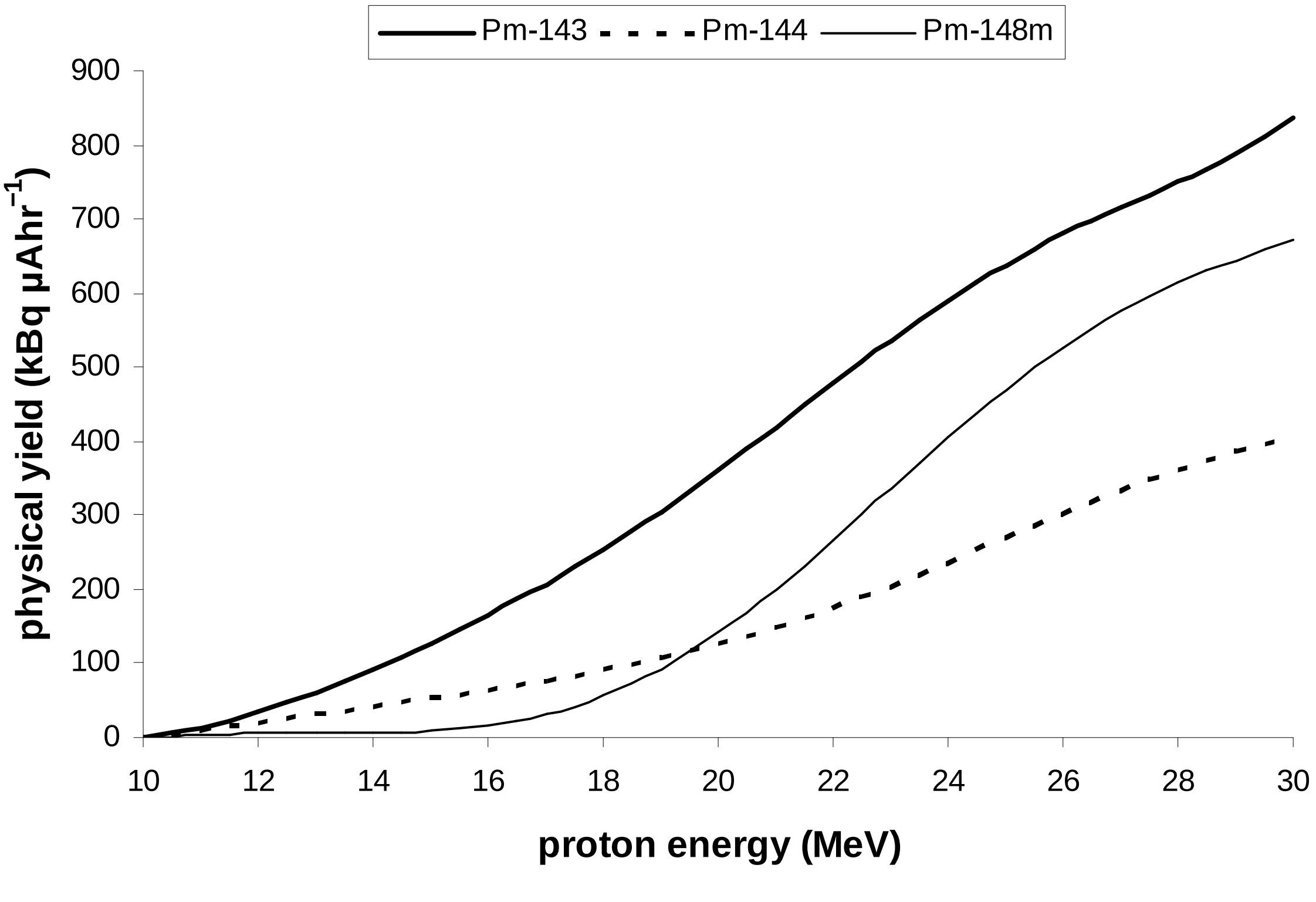

FIG. 16. Physical thick target yield of $^{143}$Pm, $^{144}$Pm, and $^{148}$Pm$^{m}$ in the $^{nat}$Nd($p$,$x$) reactions for $E_{out}$ = 10 MeV.

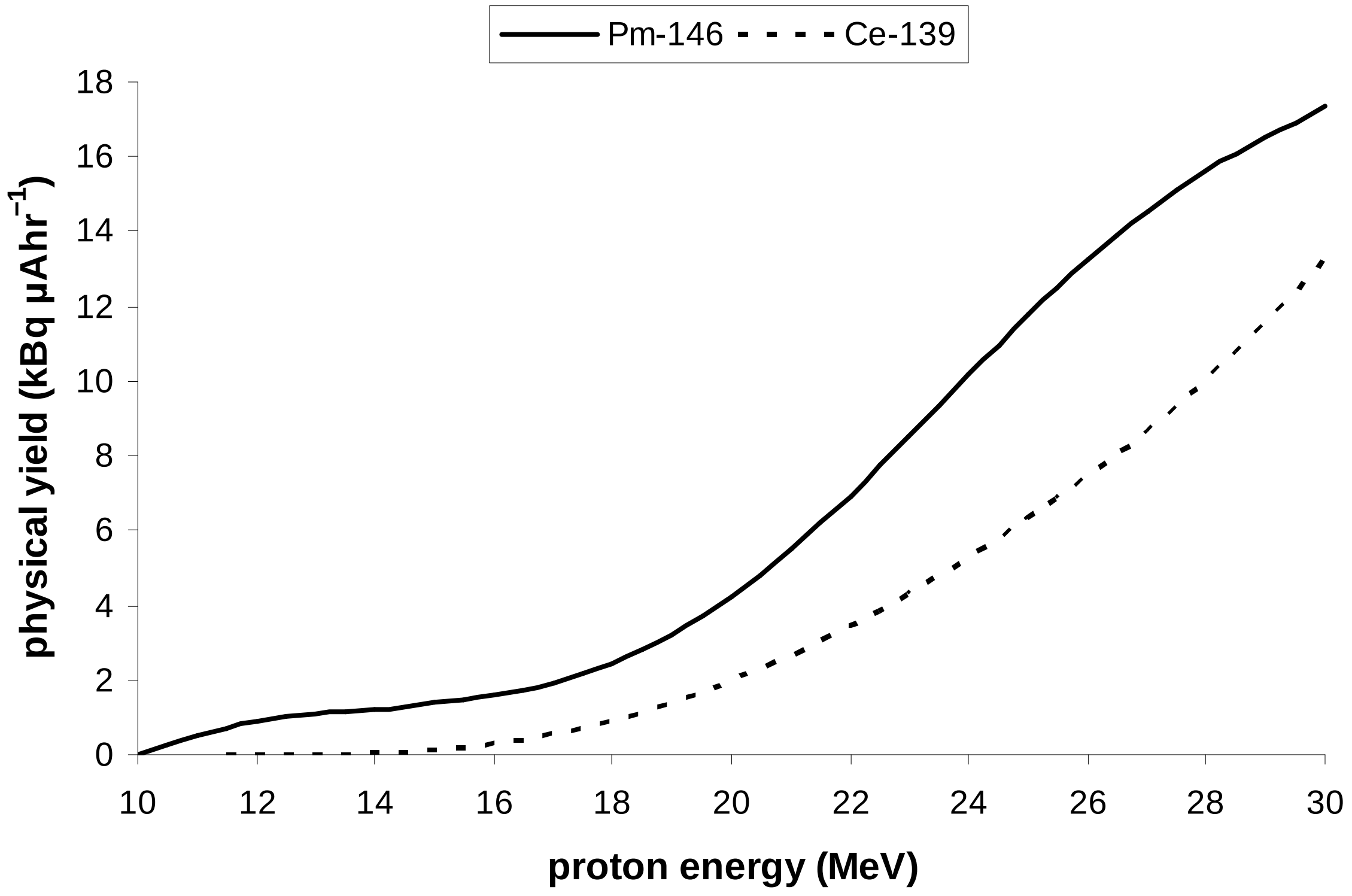

FIG. 17. Physical thick target yield of $^{148}$Pm and $^{139}$Ce in the $^{nat}$Nd($p$,$x$) reactions for $E_{out}$ = 10 MeV.